\documentclass[english,twocolumn,english,aip,reprint]{revtex4-1}
\usepackage[LGR,T1]{fontenc}
\usepackage[latin9]{inputenc}
\usepackage{array}
\usepackage{verbatim}
\usepackage{booktabs}
\usepackage{textcomp}
\usepackage{amsmath}
\usepackage{graphicx}

\makeatletter

\DeclareRobustCommand{\greektext}{%
  \fontencoding{LGR}\selectfont\def\encodingdefault{LGR}}
\DeclareRobustCommand{\textgreek}[1]{\leavevmode{\greektext #1}}
\ProvideTextCommand{\~}{LGR}[1]{\char126#1}

\providecommand{\tabularnewline}{\\}

\newcommand*{\citen}[1]{%
  \begingroup
    \romannumeral-`\x 
    \setcitestyle{numbers}%
    \cite{#1}%
  \endgroup   
}

\makeatother

\usepackage{babel}
\begin{document}

\title{Structural, optical, and electrical properties of unintentionally
doped NiO layers grown on MgO by plasma-assisted molecular beam epitaxy}

\author{Melanie Budde}

\affiliation{Paul-Drude-Institut f\"ur Festk\"orperelektronik, Leibniz-Institut im
Forschungsverbund Berlin e.V., Hausvogteiplatz 5-7, D-10117 Berlin,
Germany}

\author{Carsten Tschammer}

\affiliation{Paul-Drude-Institut f\"ur Festk\"orperelektronik, Leibniz-Institut im
Forschungsverbund Berlin e.V., Hausvogteiplatz 5-7, D-10117 Berlin,
Germany}

\author{Philipp Franz}

\affiliation{Paul-Drude-Institut f\"ur Festk\"orperelektronik, Leibniz-Institut im
Forschungsverbund Berlin e.V., Hausvogteiplatz 5-7, D-10117 Berlin,
Germany}

\author{Johannes Feldl}

\affiliation{Paul-Drude-Institut f\"ur Festk\"orperelektronik, Leibniz-Institut im
Forschungsverbund Berlin e.V., Hausvogteiplatz 5-7, D-10117 Berlin,
Germany}

\author{Manfred Ramsteiner}

\affiliation{Paul-Drude-Institut f\"ur Festk\"orperelektronik, Leibniz-Institut im
Forschungsverbund Berlin e.V., Hausvogteiplatz 5-7, D-10117 Berlin,
Germany}

\author{R\"udiger Goldhahn}

\affiliation{Institut f\"ur Experimentelle Physik, Otto-von-Guericke-Universit\"at
Magdeburg, Universit\"atsplatz 2, 39106 Magdeburg, Germany}

\author{Martin Feneberg}

\affiliation{Institut f\"ur Experimentelle Physik, Otto-von-Guericke-Universit\"at
Magdeburg, Universit\"atsplatz 2, 39106 Magdeburg, Germany}

\author{Nicolae Barsan}

\affiliation{Institut f\"ur Physikalische und Theoretische Chemie, Eberhard Karls
Universit\"at T\"ubingen, Auf der Morgenstelle 15, 72076 T\"ubingen, Germany}

\author{Alexandru Oprea}

\affiliation{Institut f\"ur Physikalische und Theoretische Chemie, Eberhard Karls
Universit\"at T\"ubingen, Auf der Morgenstelle 15, 72076 T\"ubingen, Germany}

\author{and Oliver Bierwagen}

\affiliation{Paul-Drude-Institut f\"ur Festk\"orperelektronik, Leibniz-Institut im
Forschungsverbund Berlin e.V., Hausvogteiplatz 5-7, D-10117 Berlin,
Germany}

\date{\today\\
}
\begin{abstract}
\noindent NiO layers were grown on MgO(100), MgO(110) and MgO(111)
substrates by plasma-assisted molecular beam epitaxy under Ni-flux
limited growth conditions. Single crystalline growth with a cube\nobreakdash-on\nobreakdash-cube
epitaxial relationship was confirmed by X-ray diffraction measurements
for all used growth conditions and substrates except MgO(111). A detailed
growth series on MgO(100) was prepared using substrate temperatures
ranging from 20~\textdegree C to 900~\textdegree C to investigate
the influence on the layer characteristics. Energy-dispersive X-ray
spectroscopy indicated close-to-stoichiometric layers with an oxygen
content of $\approx47$~at.\% and $\approx50$~at.\% grown under
low and high O-flux, respectively. All NiO layers had a root\nobreakdash-mean\nobreakdash-square
surface roughness below 1~nm, measured by atomic force microscopy,
except for rougher layers grown at 900~\textdegree C or using molecular
oxygen. Growth at 900~\textdegree C led to a significant diffusion
of Mg from the substrate into the film. The relative intensity of
the quasi-forbidden one\nobreakdash-phonon Raman peak is introduced
as a gauge of the crystal quality, indicating the highest layer quality
for growth at low oxygen fluxes and high growth temperature, likely
due to the resulting high adatom diffusion length during growth. Optical
and electrical properties were investigated by spectroscopic ellipsometry
and resistance measurements, respectively. All NiO layers were transparent
with an optical band gap around 3.6~eV and semi-insulating at room
temperature. However, changes upon exposure to reducing or oxidizing
gases of the resistance of a representative layer at elevated temperature
was able to confirm $p$-type conductivity, highlighting their suitability
as a model system for research on oxide-based gas sensing.
\end{abstract}
\maketitle

\section{Introduction}

\noindent Nickel oxide (NiO) is a transparent oxide due to the wide
band gap around 3.7~eV.\cite{Ohta} It crystallizes in the rock salt
crystal structure and is either unintentionally $p$\nobreakdash-type
conductive or insulating depending on the growth conditions.\cite{Rao_NiO-prop}
While highly stoichiometric NiO is considered to be insulating, hole
states induced by Ni vacancies are supposed to create the unintentional
$p$\nobreakdash-type conductivity. The $p$\nobreakdash-type conductivity
can also be created intentionally by doping with Lithium.\cite{Rao_NiO-prop,Zhang2018}
Explanations of the insulating state of NiO varied between a Mott
insulator, a charge transfer insulator, or a mixture of both.\cite{Schuler}
Furthermore, NiO is antiferromagnetic with a N\'{e}el temperature of about
525~K\cite{Schuler}. Below this temperature the crystal structure
is slightly modified into a rhombohedrally distorted form bearing
angles of > 90.1\textdegree{} instead of 90\textdegree .\cite{Massidda_Distortion}
However, for the present discussion which does not concentrate on
magnetic properties but on crystal growth, electrical and optical
properties we can safely neglect the small deviation from the perfect
cubic structure. The wide variety of properties made NiO thin films
an interesting material for many applications. First, NiO can be used
as an antiferromagnetic layer in giant magnetoresistive (GMR) spin
valve structures\cite{Hoshiya} and other magneto-electronic devices.\cite{Becker2017}
Secondly, as a \textit{p}-type transparent semiconductor it is a very
relevant material for oxide-based gas sensors\cite{Kim2014,GasSen_Overwiev}
as well as for \textit{p}\emph{n-}diodes and other transparent oxide
electronics.\cite{Ohta,Zhang_pType,Wenckstern2015} Additionally,
NiO can be used as a hole transport and electron blocking layer in
organic solar cells.\cite{Irwin} NiO films have already been grown
by many methods, including sputtering,\cite{Warot_Growth_MgO001_MgO110,Becker_NixO,NiO_Ox_states}
metal evaporation with oxygen or nitrogen dioxide inlet,\cite{Peacor}
pulsed laser deposition (PLD),\cite{Tachiki,Zhang2018} sol-gel coating,\cite{Manders_sNiO}
or plasma-assisted molecular beam epitaxy (PA\nobreakdash-MBE).\cite{Lind}
As an alloy together with MgO it is also interesting for deep\nobreakdash-ultraviolet
photodetectors, offering a band gap tuning between 3.6 and 7.8~eV.\cite{Mares_NiMgO}
\\

\noindent To exclude the disturbing effects of grain boundaries for
basic investigations of the material single crystalline NiO layers
are necessary. These layers can serve as model system with reduced
complexity to investigate fundamentals of NiO-based applications,
such as oxide-based gas-sensing.\cite{Rombach2016} The common rock-salt
crystal structure with similar lattice constants of MgO (0.4212~nm)
and NiO (0.4176~nm) and low lattice mismatch <~1~\% makes MgO a
widely-used, suitable substrate to epitaxially grow high quality NiO
layers.\cite{Lind,Zhang2018} In addition, MgO was found to be a good
electron blocking layer for GaN/NiO based diodes.\cite{Wang_NiO/MgO/GaN_Diode}
Detailed investigations about the epitaxial growth of NiO at different
growth conditions, however, are very rare: Warot et al. studied the
growth of NiO by PLD on different orientations of MgO.\cite{Warot_Growth_MgO001_MgO110,Warot_Islands_MgO111,Warot_Morph_MgO110,Warot_Morpho_all_Orientations}
Lind et al. observed single crystalline NiO films on MgO(100) by PA-MBE
only for growth temperatures up to 260\textdegree C and polycrystalline
growth for higher temperatures.\cite{Lind} \\

\noindent In this study we investigate the PA\nobreakdash-MBE growth
and properties of NiO thin films on different MgO orientations and
under a wider range of growth conditions with the goal of realizing
high quality single crystalline layers. Furthermore, we propose a
new metrics to measure the crystalline quality of NiO using Raman
spectroscopy. For all properties a reference NiO bulk sample was measured.

\section{Experiment}

\noindent For the MBE growth, quarters of 2-inch MgO(100), MgO(110)
and MgO(111) substrates (from CrysTec GmbH) were used. For layers
grown on MgO(100) growth temperature and oxygen flux were varied with
the idea to modify the concentration of Ni vacancies. The growth temperatures
are defined by the substrate heater temperatures, measured by a thermocouple
between substrate and heating filament. A 1~\textmu m thick layer
of sputter-deposited titanium layer on the rough backside of the substrate
improved the substrate heating by absorbing the radiation from the
heating filament. Activated oxygen was provided by flux passing a
controlled molecular oxygen through an RF plasma source directed to
the substrate. Before growth, the surface quality of all substrates
was improved by a plasma treatment in the growth chamber at various
conditions defined by temperatures ranging from 700~\textdegree C
to 900~\textdegree C, oxygen fluxes ranging from 0.3 to 3~standard
cubic centimeters per minute (sccm) at plasma powers ranging from
150~W to 300~W for 20 to 30 minutes. This treatment helps to reduce
organic contaminations at the surface and increase the surface crystalline
order. For the growth, an oxygen plasma and a nickel effusion cell
were used. As a protection of the nickel effusion cell, the temperature
was kept well below the melting point of nickel (1455~\textdegree C),
at 1380~\textdegree C. The beam equivalent pressure (BEP), which
is proportional to the particle flux, was measured by a nude filament
ion gauge positioned at the substrate location and removed before
growth. The resulting nickel BEP was between $7.7\text{\ensuremath{\cdot}}10^{-9}$~mbar
and $1.3\cdot10^{-8}$~mbar, which led to growth rates between 0.05~\r{A}/s
and 0.09~\r{A}/s. The growth temperature for layers grown on MgO(100)
was varied between 20~\textdegree C and 900~\textdegree C. Furthermore,
different activated oxygen fluxes were used: series one (S1) was grown
with a high oxygen flux of 1~sccm at a plasma power of 300~W with
resulting oxygen BEP of $\approx1\cdot10^{-5}$~mbar. Series two
(S2) was grown with a strongly reduced active oxygen flux of 0.3~sccm
at 150~W with resulting oxygen BEP of $\approx3\cdot10^{-6}$~mbar.
Higher oxygen fluxes, which should be beneficial for the formation
of Ni vacancies, led to the problem of NiO formation at the orifice
of the nickel cell, which continuously reduces the Ni flux and can
eventually lead to the complete closure of the Ni cell. In the following
the samples will be named by their series and growth temperature,
for example the sample from series 1 grown at 700~\textdegree C is
called S1-700. The layers on MgO(110) and MgO(111) were grown at 700~\textdegree C
with an oxygen flux of 1~sccm and a power of 300~W. Besides the
plasma assisted growth, one layer was grown at 700~\textdegree C
on MgO(100) with molecular oxygen and a flux of 0.3~sccm.\\
\\
\begin{figure}
\noindent \begin{centering}
\includegraphics[width=7.5cm]{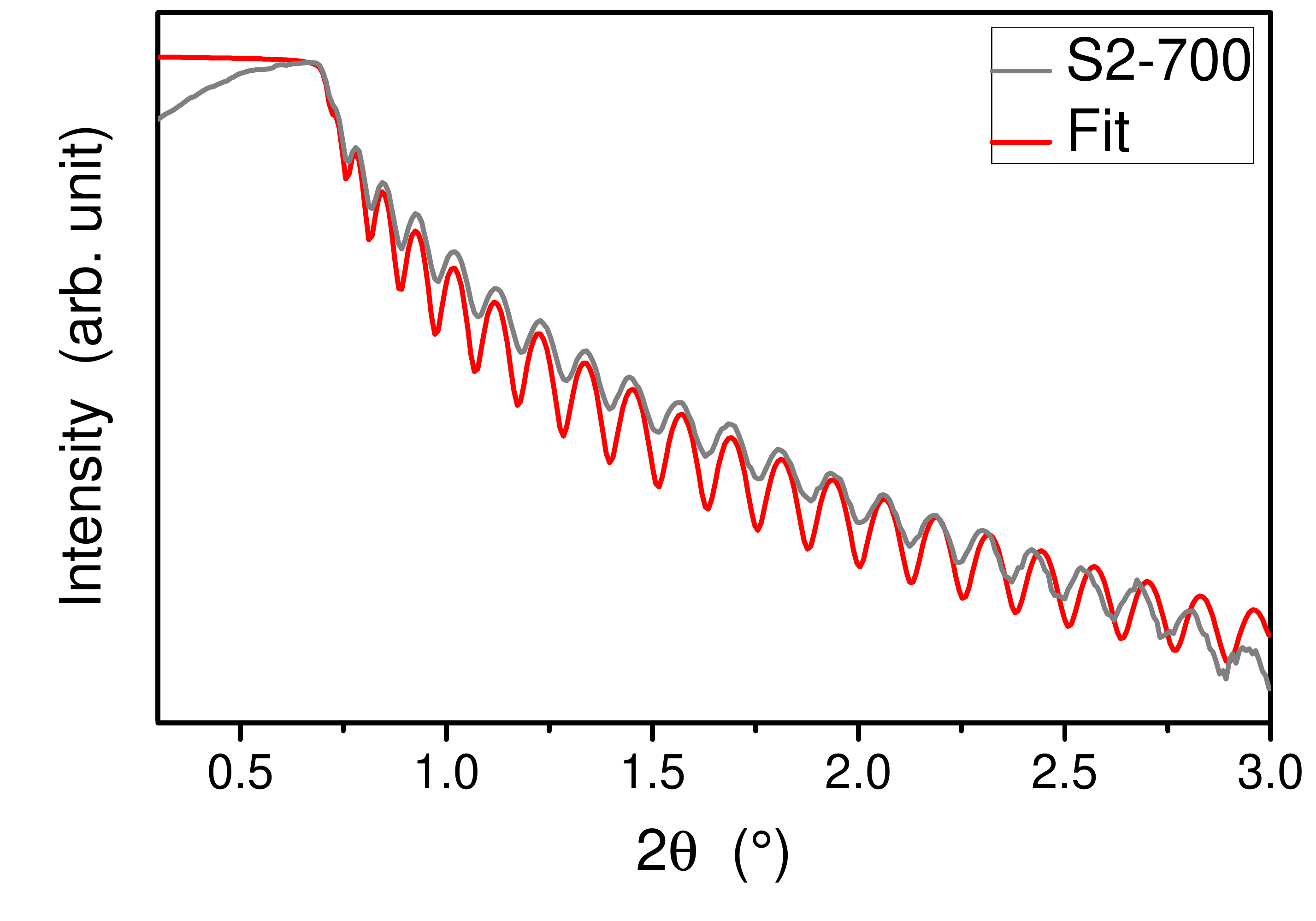}
\par\end{centering}
\caption{XRR curve and fit of S2-700, as an example of thickness determination.
\label{fig:XRR-curves}}
\end{figure}
After the growth, all layers were investigated by different ex\nobreakdash-situ
methods. First, the layer thickness was determined by fitting measured
X\nobreakdash-ray reflectometry (XRR) curves assuming only the NiO
layer on top of the MgO substrate. A representative XRR curve and
its corresponding fit curve are shown in Fig.~\ref{fig:XRR-curves}
for S2-700. The extracted thicknesses of this layer and all other
layers grown with an oxygen plasma on MgO(100) are shown in Tab.~\ref{tab:Thickness}.
The extracted layer thickness of the sample grown with molecular oxygen
is 39~nm. The extracted layer thickness on MgO(110) and MgO(111)
are 30~nm and 35~nm, respectively. Secondly, different X\nobreakdash-ray
diffraction (XRD) scans were used to determine the epitaxial relationship
between substrate and layer.
\begin{table}
\noindent \centering{}\caption{Measured thicknesses for all layers grown on MgO(100) with PA-MBE,
except the 900~\textdegree C. This sample was not measurable by XRR.
\label{tab:Thickness}}
\begin{tabular*}{8cm}{@{\extracolsep{\fill}}>{\centering}b{1.3cm}>{\raggedright}m{2cm}>{\centering}m{0.9cm}>{\centering}m{0.9cm}>{\centering}m{0.9cm}>{\centering}m{0.9cm}}
\toprule 
\textbf{S1} & \textbf{$\mathbf{T}_{\mathbf{g}}$ {[}\textdegree C{]}} &  & \textbf{250} & \textbf{450} & \textbf{700}\tabularnewline
\cmidrule{2-6} 
 & \textbf{$\mathbf{d}$ {[}nm{]}} &  & 50 & 24 & 51\tabularnewline
\midrule 
\textbf{S2} & \textbf{$\mathbf{T}_{\mathbf{g}}$ {[}\textdegree C{]}} & \textbf{20} & \textbf{200} & \textbf{400} & \textbf{700}\tabularnewline
\cmidrule{2-6} 
 & \textbf{$\mathbf{d}$ {[}nm{]}} & 62 & 60 & 53 & 65\tabularnewline
\bottomrule
\end{tabular*}
\end{table}

\noindent Further layer properties were investigated for the growth
series on MgO(100). The effect of different growth parameters on the
surface morphology was analyzed by atomic force microscopy (AFM) in
the peak force tapping mode using a Bruker \textquotedblleft Dimension
edge\textquotedblright{} with the ``ScanAsyst'' technology. Scanning
electron microscope-based energy dispersive X\nobreakdash-ray spectroscopy
(EDX) with a Zeiss ULTRA 55 was used to determine the proportion of
nickel and oxygen for selected samples, using an electron energy of
3~keV and a beam current of about 1~nA. The X-ray radiation is measured
and analyzed by an EDAX system with a SDD Apollo XV detector. Raman
spectroscopy was utilized to assess the NiO crystal quality. The spectra
were recorded at room temperature in backscattering geometry using
the 325-nm line of a Cd-He ion laser and a 405\nobreakdash-nm diode
laser for optical excitation. In addition, optical characteristics
were determined by spectroscopic ellipsometry measurements in the
range of 1.2 to 6.5~eV for three angles of incidence (50\textdegree ,
60\textdegree , 70\textdegree ) to improve the reliability of the
obtained dielectric functions (DFs). The DFs were obtained by multilayer
modeling of the ellipsometric data\cite{DF_Goldhahn}, where the surface
roughness was modeled using the Bruggeman effective medium approximation.\cite{Bruggeman}
A similar procedure has been, for example, used for the investigation
of the cubic In$_{2}$O$_{3}$.\cite{Feneberg_In2O3} As Hall measurements
failed due to the high resistivity of the samples, the electrical
characteristics were investigated using interdigitated contact patterns
(idcs) to reduce the measured resistance with respect to the sheet
resistance. The idcs consist of a layer of 20~nm Pt covered by 150~nm
Au for S1. For S2 we reduced the metal stack to 15~nm Pt and 80~nm
Au to simplify the lift-off process. Room temperature current-voltage
(I-V) characteristics of these patterns were measured to estimate
the sheet resistance. Furthermore, the change of resistance of a heated
sample during exposure to oxidizing and reducing gases was used to
confirm its \textit{p}-conductivity.\cite{Barsan_gasssensing,Kim2014} 

\section{Results and discussion}

\subsection{Growth temperature\label{subsec:Growth-temperature}}

\begin{figure}
\begin{centering}
\includegraphics[width=8.5cm]{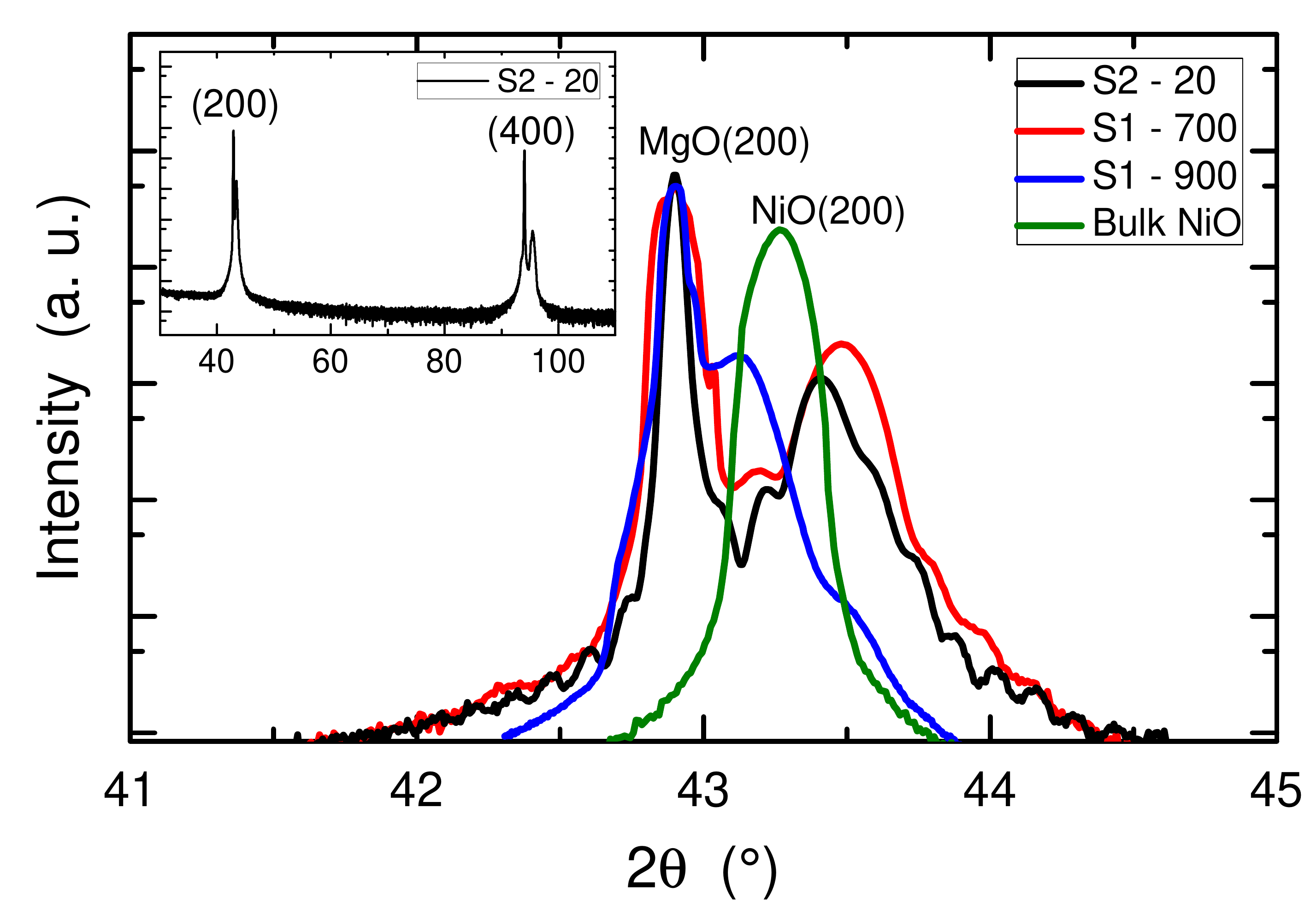}
\par\end{centering}
\caption{$2\theta-\omega$ scans of three NiO layers at different growth temperatures
and the NiO bulk reference sample. A clear difference is visible between
the sample grown at 900~\textdegree C and samples from lower growth
temperatures. The inset shows the scan of sample S2-20 for an extended
range, showing no additional peaks, which confirms the monocrystalline
behavior.\label{fig:2=0003B8=002011=0003C9-scan-MgO100}}
\end{figure}

\noindent The out\nobreakdash-of\nobreakdash-plane orientation of
layers grown on MgO(100) at different growth temperatures was investigated
by symmetric $2\theta-\omega$ scans. Fig.~\ref{fig:2=0003B8=002011=0003C9-scan-MgO100}
shows representative results for different layers as well as the bulk
reference sample. For all grown layers only the NiO(200) and the MgO
(200) substrate peak (and higher diffraction orders of the (100) planes
as shown in the inset) are present, indicating exclusively (100)-oriented
films. This is in contrast to the observations by Lind et al. who
could only grow single-crystalline layers up to 250~\textdegree C
- 260~\textdegree C.\cite{Lind} However, much higher growth rates
(2.0\nobreakdash-2.5~\r{A}/s) were used in that case. Thus, a low growth
rate may be important for the high temperature growth of single crystalline
NiO.

\noindent ``Pendell\"osung'' thickness fringes next to the NiO(200)
peak are present for all growth temperatures, arising from the interference
between the X-rays reflected at the two interfaces (air\nobreakdash-layer,
layer\nobreakdash-substrate) for layers with regular lattice periodicity.\cite{Fewster_Fringes}
The bulk reference sample only shows the NiO(200) peak without fringes
due to the missing substrate.

\noindent The exact position of the NiO(200) diffraction peak is related
to the film lattice constant in the growth direction. Strained growth
of NiO on MgO is expected up to a critical thickness of about 60~nm,\cite{James_Strain}
due to the lattice mismatch. For pseudomorphic growth on MgO, the
adaption of the in\nobreakdash-plane direction to the higher MgO
lattice constant would result in compressive out\nobreakdash-of\nobreakdash-plane
strain. This leads to a lower out\nobreakdash-of\nobreakdash-plane
lattice constant and therefore to a higher $2\theta$ angle for the
(200) peak compared to unstrained NiO, which can be seen for S1-700
and S2-20 in Fig.~\ref{fig:2=0003B8=002011=0003C9-scan-MgO100}.
For NiO on MgO an in-plane strain of 0.00833 can be calculated by
the relaxed lattice constants assuming pseudomorphic growth. Using
the Poisson ratio from James et al. of 0.21,\cite{James_Strain} an
out-of-plane strain ($\epsilon_{out}$) of about 0.00443 should be
measured. Similar values could be found for our layers deduced from
the peak positions in the $2\theta-\omega$ scans (0.00417 - 0.00451),
but some layers already started to relax due to their higher layer
thickness, as it can be seen for the S2\nobreakdash-20 sample in
Fig.~\ref{fig:2=0003B8=002011=0003C9-scan-MgO100} ($\epsilon_{out}$=0.00324).

\noindent Interestingly, the growth at 900~\textdegree C led to a
NiO(200) peak at lower $2\theta$ than the bulk value (see Fig.~\ref{fig:2=0003B8=002011=0003C9-scan-MgO100}),
corresponding to a higher out-of-plane lattice constant than that
of relaxed bulk NiO. This behavior cannot be explained by strain from
the substrate. Instead, a higher lattice constant up to that of MgO
can be realized in the alloy Ni$_{1-x}$Mg$_{x}$O.\cite{Boutwell_NiMgO}
Using the lattice constant vs. Mg-content data of Ref.~{[}\citen{Boutwell_NiMgO}{]}
a Mg concentration of x=46~\% can be estimated using the lattice
constant of 0.419~nm deduced from the $2\theta$ peak position of
S1-900. 
Thus, the measured (200) peak position shift indicates significant
diffusion of Mg from the substrate into the layer activated by the
high growth temperature, which would also explain the missing XRR
oscillations as diffusion leads to a smeared-out density profile at
the substrate-film interface. However, a thickness of about 40~nm
still was measurable by the weak Pendell\"osung fringes in the $2\theta-\omega$
scan around the (200) peak. Mg-diffusion into the layer is corroborated
by the presence of a Mg-peak in SEM-EDX measurement for the S1-900
sample at an electron beam energy of only 2~keV, which excludes penetration
into the MgO substrate. 

\noindent Hence, epitaxial NiO(100) layers can be grown on MgO(100)
at growth temperatures in the range of 20\textdegree C to 700\textdegree C,
whereas a higher temperature of 900\textdegree C results in significant
Mg incorporation from the substrate into the layer. 

\subsection{Epitaxial relationship }

\subsubsection*{MgO(100) substrate}

\noindent 
\begin{figure}
\begin{centering}
\includegraphics[width=7cm]{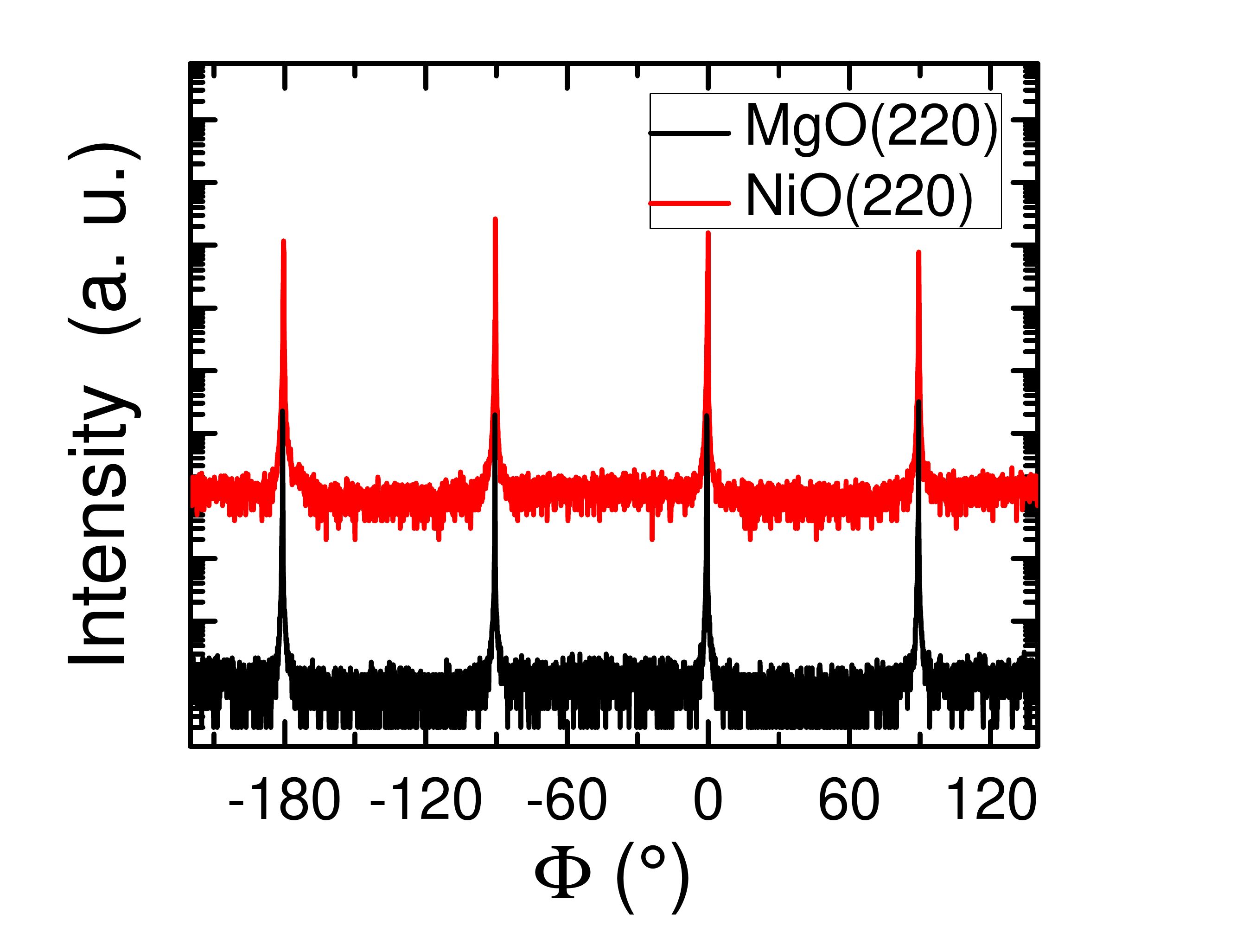}
\par\end{centering}
\caption{An example of a $\Phi$-scan from a NiO layer grown on MgO(100). As
all samples, the layer from S2\protect\nobreakdash-20 shows a fourfold
symmetry and a cube-on-cube growth. \label{fig:Phi-MgO100}}
\end{figure}

\noindent To determine the in-plane expitaxial relationship and the
potential existence of rotational domains, XRD peaks of inclined lattice
planes were measured by $\Phi$-scans with rotational angle $\Phi$
around the surface normal. These measurements were performed in the
skew-symmetric geometry with the sample tilted by the angle $\Psi$
between the surface normal and the normal of the inclined lattice
planes. Fig.~\ref{fig:Phi-MgO100} shows such a $\Phi$-scan of the
NiO(220) and MgO(220) peak of sample S2\nobreakdash-20. It indicates
a fourfold rotational symmetry for substrate and layer, as expected
for the (100)-oriented structure, confirming the absence of rotational
domains in substrate and film. In addition, the peaks of the substrate
are at the same $\Phi$ positions as the layer peaks, indicating coinciding
$\left\langle 110\right\rangle $ directions of substrate and layer.
Therefore, the epitaxial relationship is cube\nobreakdash-on\nobreakdash-cube,
i.e.: 
\[
NiO[100]\:||\:MgO[100]
\]

\noindent 
\[
NiO[010]\:||\:MgO[010]
\]
This cube\nobreakdash-on\nobreakdash-cube growth is maintained under
all investigated growth conditions, including the growth with molecular
oxygen and that at high growth temperature with Mg-incorporation,
as confirmed by XRD characterization (not shown) of all samples.\\

\noindent Rocking curves (\textgreek{w}\nobreakdash-scan) measured
by XRD are normally used for quality investigations of thin layers.
The full\nobreakdash-width\nobreakdash-at\nobreakdash-half\nobreakdash-maximum
(FWHM) of the peak gives information about the crystal perfection,
since it correlates with the range of tilt angles of the measured
lattice planes. Higher FWHM can be created by defects, such as misfit
dislocations,\cite{Vickers_X-ray_nitrides} hence a smaller FWHM indicates
a higher quality of the crystal. The FWHM can, however, also be affected
by curvature or substrate quality. Indeed, MgO substrates often consist
of multiple macroscopic domains with low-angle domain boundaries and
a wide range of quality can occur even for the same vendor and batch.\cite{Schroeder_Poor_MgO}
Depending on the size, tilt and occurence of those macroscopic domains,
a MgO peak can consist of one or multiple features,\cite{Schroeder_Poor_MgO}
which can be seen by closely inspecting the MgO(200) peak in Fig.~\ref{fig:2=0003B8=002011=0003C9-scan-MgO100}
showing a sharp peak for S2-20 but multi-peak structure for S1-700
and S1-900. The FWHMs of the layers grown on MgO(100) are similar
for most samples and around 0.05\textdegree{} to 0.07\textdegree .
Only series two shows higher values (0.21~\textdegree -0.42~\textdegree )
for all temperatures, except 20~\textdegree C.
\begin{figure}
\noindent \begin{centering}
\includegraphics[width=8.5cm]{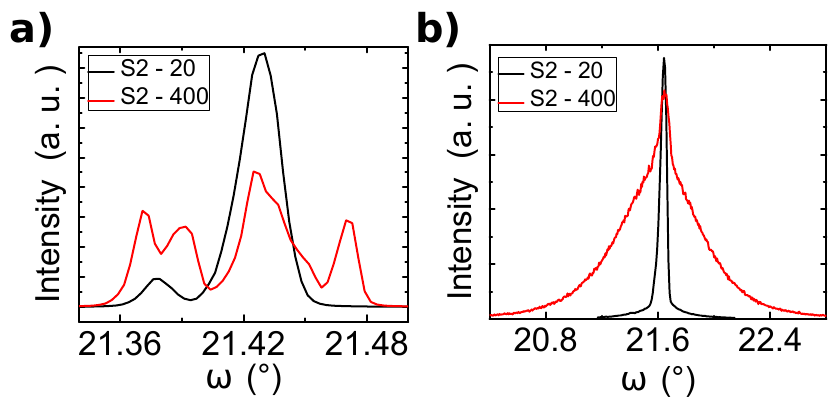}
\par\end{centering}
\caption{a) Rocking curves of the substrates of S2-20 and S2-400 plotted with
a linear intensity scale, indicating a better substrate quality for
S2-20. The resulting rocking curves observed for the NiO layers are
shown in b). The measured intensity of S2-400 is multiplied by 5 for
a better comparison with S2-20. The multiple peaks of the substrate
of S2-400 result in a broader peak for the NiO layer, too. \label{fig:FWHM}}
\end{figure}
An example of the two different rocking curves is shown in Fig.~\ref{fig:FWHM}.
The broader peaks can be explained by the quality of the substrate
which depends on the orientation of the different macroscopic domains
it consists of. Due to the strong impact of the fluctuating substrate
quality, no correlation between the quality of the layers and the
temperature or oxygen flux influence can be extracted from the FWHM.
However, the measured values can be considered as an upper bound estimate
with low FWHM (<0.07\textdegree ) on the higher quality substrates.
\begin{figure}
\noindent \begin{centering}
\includegraphics[width=8.5cm]{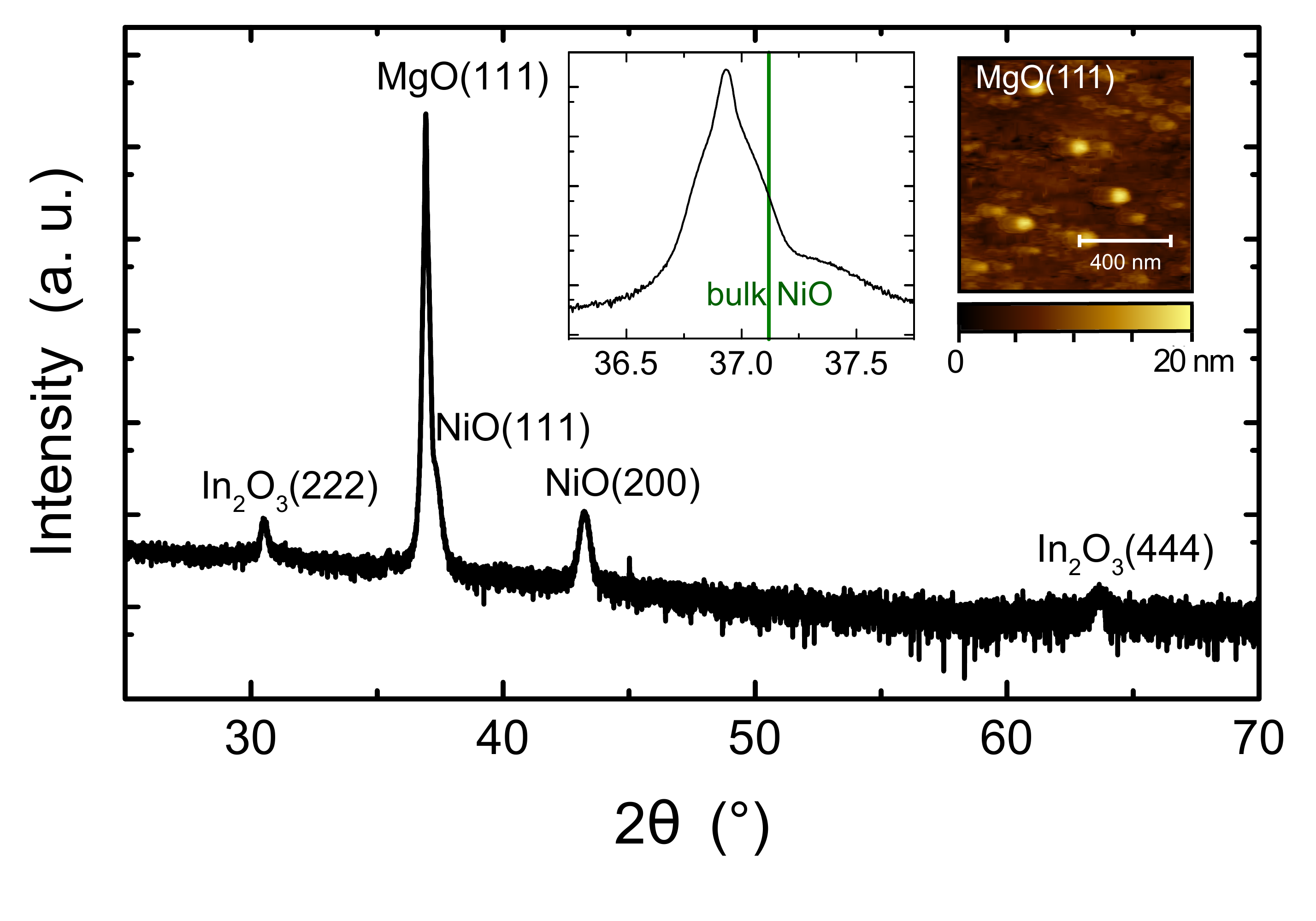}
\par\end{centering}
\caption{$2\theta-\omega$ scan of the NiO layer grown on MgO(111). The full
scan shows additional indium oxide peaks, because indium bonding was
used during MBE growth. Besides the substrate peak, (111)- and (100)-oriented
NiO crystals are measured by XRD. The NiO(111) peak is near the position
of the MgO(111) peak, but is clearly visible in the inset for angles
from 36\textdegree{} to 38\textdegree . The AFM image shows the low
surface quality of a MgO(111) substrate.\label{fig:2tw_MgO111}}
\end{figure}

\subsubsection*{MgO(111) substrate}

\noindent The $2\theta-\omega$\-\- scan of the layer grown on MgO(111)
is shown in Fig.~\ref{fig:2tw_MgO111}. It indicates the presence
of a NiO(200) and a NiO(111) peak (seen as a shoulder at the high
angle-side of the MgO(111) peak from the substrate). The NiO(111)
peak can be clearly distinguished from the MgO(111) in the inset of
Fig.~\ref{fig:2tw_MgO111}. Therefore, a single crystalline growth
on the MgO(111) substrate did not take place under our growth conditions,
likely related to a low surface quality of the substrate (see AFM
image in Fig.~\ref{fig:2tw_MgO111}) with a RMS roughness of 2.3~nm,
that may trigger the formation of the thermodynamically stable (100)
surface. The higher stability of the \{100\}- compared to the \{111\}-
facet has been often observed by the growth of NiO tetrahedrons with
\{100\} planes on MgO(111) and Al$_{2}$O$_{3}$(0001) substrates.\cite{Warot_Morpho_all_Orientations,Warot_Islands_MgO111,Becker2017}
Additionally, a weak indium oxide peak is visible which is due to
oxidized traces of indium that unintentionally covered small parts
of the surface during In-bonding of the substrate. \\
\begin{figure}
\noindent \begin{centering}
\includegraphics[width=8cm]{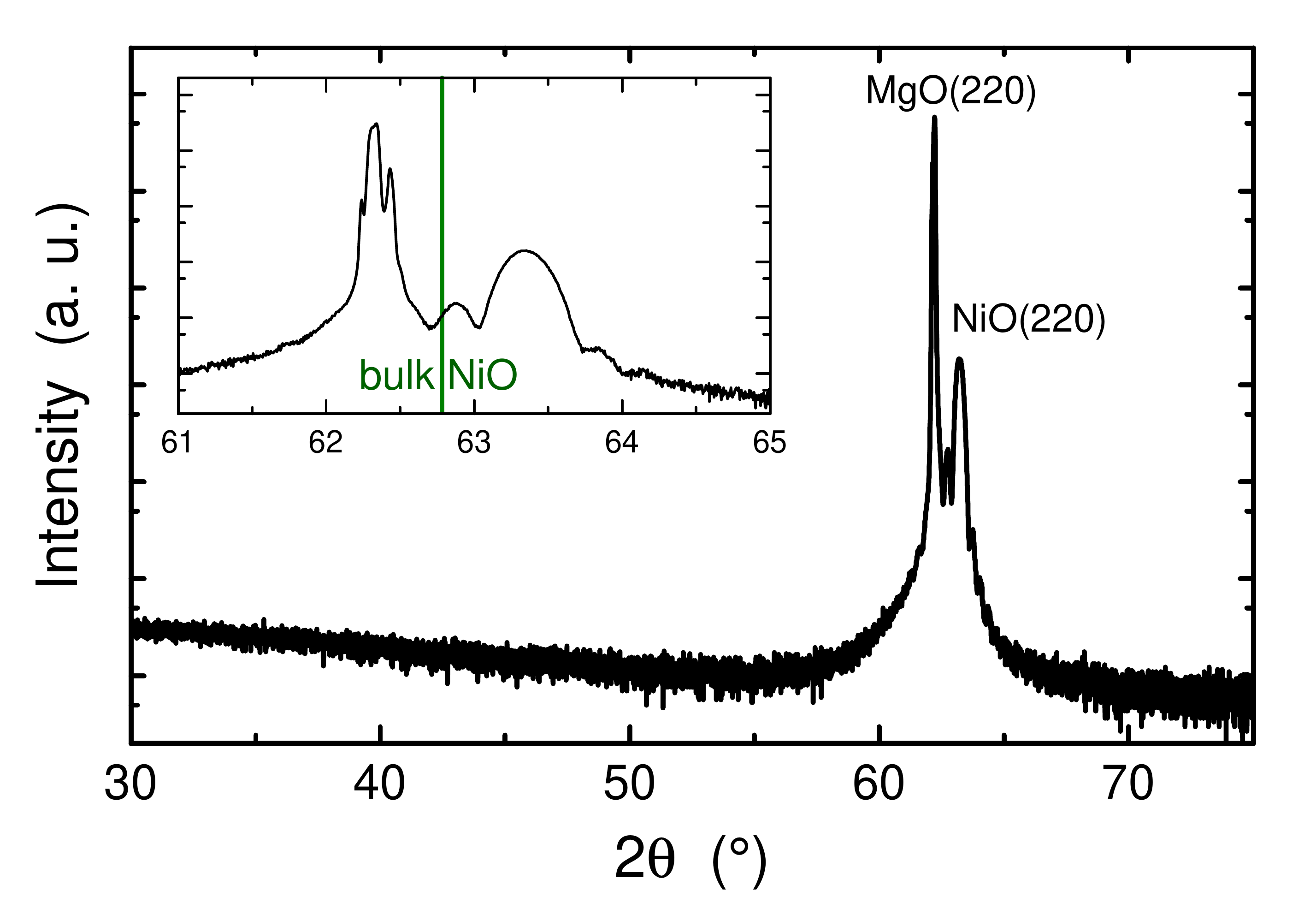}
\par\end{centering}
\caption{$2\theta-\omega$ scan of the MgO(110) sample. The NiO layer is stressed
and (110)-oriented. \label{fig:2tw MgO110}}
\end{figure}

\subsubsection*{MgO(110) substrate}

\begin{figure}
\noindent \begin{centering}
\includegraphics[width=7cm]{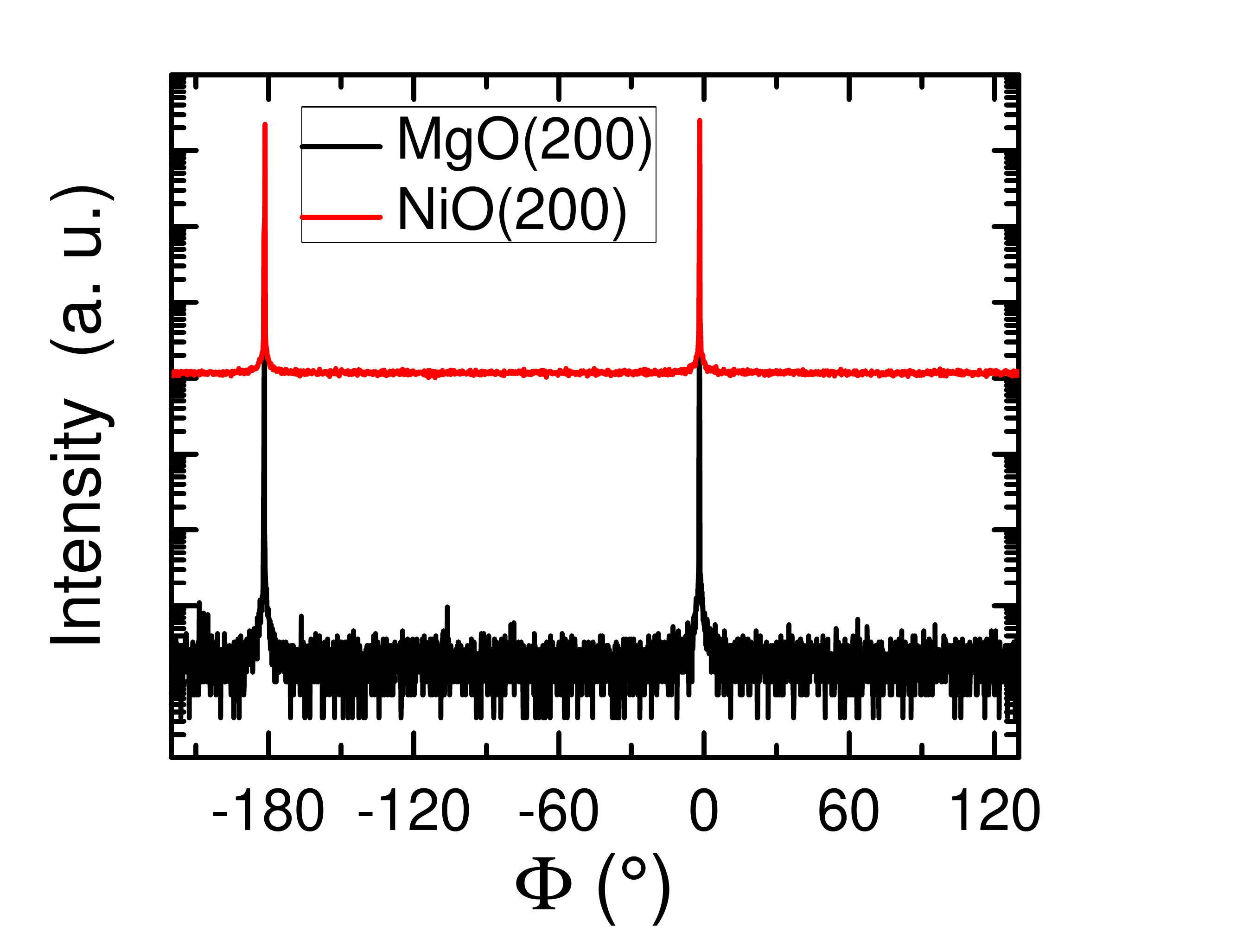}
\par\end{centering}
\caption{$\Phi$-scan of a NiO layer on MgO(110) showing a twofold symmetry.
The peaks of the NiO(200) and MgO(200) are at the same angle, which
describes a cube-on-cube growth. \label{fig:Phi_MgO(110)}}
\end{figure}

\noindent In the $2\theta-\omega$ scan of the MgO(110) sample a NiO(220)
peak is measured beside the substrate peak (see Fig.~\ref{fig:2tw MgO110}),
indicating a single crystalline layer. For this layer the $\Phi$
scan of the NiO(200) reflex is shown in Fig.~\ref{fig:Phi_MgO(110)}.
It shows a twofold symmetry, which is expected for the (110) surface
orientation. In addition, the peaks of substrate and layer are at
the same $\Phi$ positions. Thus, the epitaxial relationship is cube\nobreakdash-on\nobreakdash-cube,
i.e:

\noindent 
\[
NiO[1\text{10}]\:||\:MgO[110]
\]

\noindent 
\[
NiO[1\overline{1}0]\:||\:MgO[1\overline{1}0]
\]

\noindent 

\subsection{Oxygen-rich growth and oxygen stoichiometry of the layer}

\noindent Generally, the growth rate of an oxide film can be limited
by the metal flux, corresponding to O-rich growth conditions, or by
the O-flux, corresponding to metal-rich growth conditions.\cite{Vogt2016a}
All films were grown at the same Ni cell temperature, resulting in
an approximately constant Ni-flux. The measured growth rates ranging
from 0.05~\r{A}/s to 0.09~\r{A}/s did not show a systematic dependence on
the reactive O-flux, despite a strong variation of the oxidizing agent
ranging from a low flux of 0.3~sccm molecular oxygen to a high flux
of 1~sccm of plasma-activated oxygen. The growth rate rather decreased
over time due to the gradual formation of NiO at the orfice of the
Ni cell. Consequently, the growth was metal-flux limited, O-rich under
all investigated growth conditions.

\noindent Under different growth conditions many stoichiometries of
nickel oxide have been prepared by other growth techniques: Oxygen-rich
stoichiometries of sputtered and electro-deposited Ni$_{3}$O$_{4}$,
Ni$_{2}$O$_{3}$ and NiO$_{2}$\cite{NiO_Ox_states} as well as ion-beam
sputtered Ni$_{x}$O ranging from slightly O-rich (x=0.9) to Ni-rich
(x=1.45)\cite{Becker_NixO} stoichiometries have been reported. Interestingly,
XRD of the latter ones showed a cubic-NiO peak that was almost unchanged
with stoichiometry.\cite{Becker_NixO} XRD of all our layers indicates
only the cubic phase by the presence of the NiO(200) and (400) peaks
and the peak seen in the $\Phi$-scans. A small shift of the NiO(200)
peak of S2-700 with respect to S1-700 is visible but is related rather
to strain than to stoichiometry as the thickness of S1-700 and S2-700
is below and above the critical thickness of 60~nm, respectively.

\noindent To clarify the effect of the different oxygen fluxes during
growth on the stoichiometry of our films further investigations by
EDX were performed with the samples S1\nobreakdash-250, S1\nobreakdash-700,
and S2\nobreakdash-700. The results are shown in Tab.~\ref{tab:EDX relation}.
Due to the accuracy of 1\nobreakdash-2~\% of the measurement both
samples of S1 can be seen as stoichiometric NiO, whereas a small Ni
excess of <5~At\% was found in S2\nobreakdash-700. This Ni excess
would be consistent with the lower O-flux used in S2 and could manifest
itself in the existence of Ni-interstitials or O-vacancies.

\noindent Hence, in our study both oxygen fluxes lead to nearly stoichiometric
NiO films grown under O-rich conditions. 
\begin{table}
\noindent \centering{}\caption{Oxygen and nickel proportion obtained by EDX measurements of selected
NiO layers grown on MgO(100). \label{tab:EDX relation}}
\begin{tabular*}{8cm}{@{\extracolsep{\fill}}>{\centering}m{1.3cm}>{\raggedright}m{2cm}>{\centering}m{1.2cm}>{\centering}m{1.2cm}}
\toprule 
 & \textbf{$\mathbf{T}_{\mathbf{g}}$ {[}\textdegree C{]}} & \textbf{250} & \textbf{700}\tabularnewline
\cmidrule{2-4} 
\textbf{S1} & \textbf{At\% O} & 49.9 & 49.3\tabularnewline
\cmidrule{2-4} 
 & \textbf{At\% Ni} & 50.1 & 50.7\tabularnewline
\midrule 
 & \textbf{$\mathbf{T}_{\mathbf{g}}$ {[}\textdegree C{]}} &  & \textbf{700}\tabularnewline
\cmidrule{2-4} 
\textbf{S2} & \textbf{At\% O} &  & 47.2\tabularnewline
\cmidrule{2-4} 
 & \textbf{At\% Ni} &  & 52.8\tabularnewline
\bottomrule
\end{tabular*}
\end{table}

\subsection{Surface morphology and annealing}

\begin{figure}
\begin{centering}
\includegraphics[width=7.5cm]{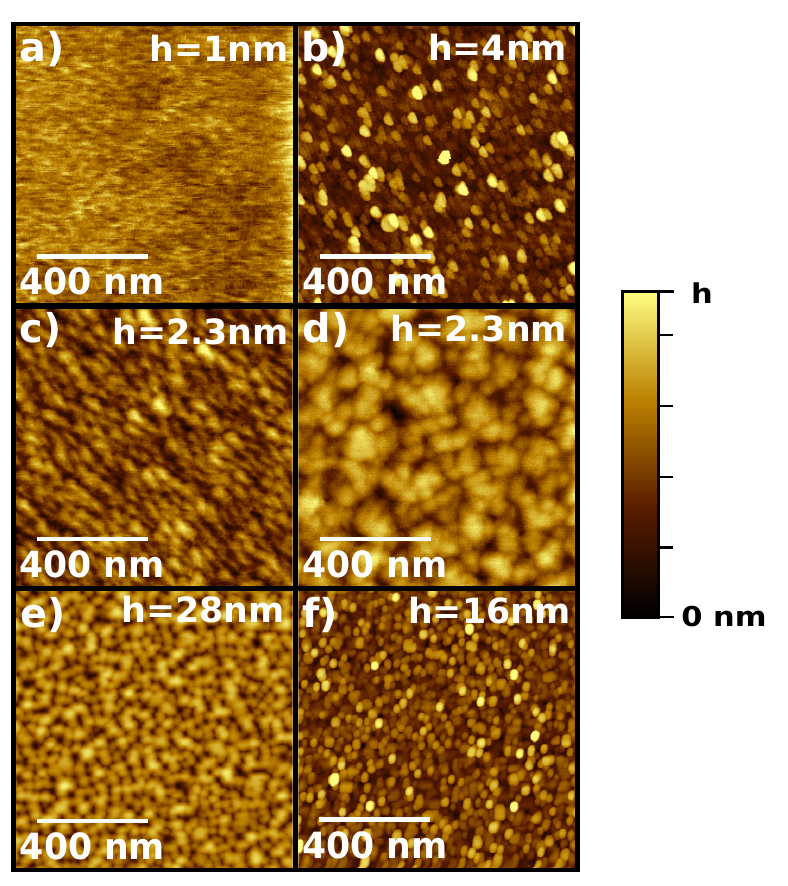}
\par\end{centering}
\caption{AFM images of the MgO(100) substrate (a), the S2\protect\nobreakdash-20
NiO layer (b), the S1\protect\nobreakdash-700 layer (c) and the S2-700
layer (d). All show a surface composed of grains. However,the size
is smaller for the lower temperature and for the first series at 700~\textdegree C.
The AFM image of the Mg-incorporated S1-900 sample is shown in (e)
and the layer grown with molecular oxygen in (f). The AFM image shows
a high roughness for both.\label{fig:AFM_MgO100}}
\end{figure}

\noindent The morphology of all samples on MgO(100) was investigated
by AFM. The clean MgO substrate is shown in Fig.~\ref{fig:AFM_MgO100}~a).
The surface is flat and has a root-mean-squared (RMS) roughness of
about 0.1~nm. Furthermore, AFM images of S2-20, S1-700 and S2-700
are also shown in Fig.~\ref{fig:AFM_MgO100}(b), (c) and (d), respectively.
All layers show a surface consisting of grains, independent of temperature
and oxygen flux. As it is shown in Tab.~\ref{tab:RMS+GrainSize},
a correlation between growth temperature and grain size can be seen.
Overall, the sizes of series two are bigger, indicating a higher adatom
surface diffusion length consistent with the lower O-flux and a higher
temperature. In addition, the table shows the RMS roughness of all
samples. However, no clear relation between the temperature and the
roughness can be found. Solely, the roughness of S1-900 (with Mg incorporated
from the substrate) of 3.7~nm is significantly higher than that of
NiO grown at lower temperature, although the average grain size is
still around 50~nm (see Fig.~\ref{fig:AFM_MgO100} e).
The layer grown with molecular oxygen (RF plasma power 0~W) also
shows a higher roughness of about 2.5~nm due to the formation of
tall islands (Fig.~\ref{fig:AFM_MgO100} f). This indicates a low
surface diffusion length even at 700~\textdegree C, presumably due
to the reduced adatom surface diffusion length by the lower oxidation
power of molecular oxygen. This is in agreement with the investigation
by Peacor et al. who observed NiO island formation for growth with
molecular oxygen and smoother surfaces for samples grown with the
more reactive NO$_{2}$.\cite{Peacor} For growth temperatures higher
than 400~\textdegree C even unoxidized nickel has been observed in
O$_{2}$ grown layers. Peacor et al. further indicate a correlation
of the small grain size with a low surface diffusion rate for temperatures
under 300~\textdegree C and insufficient oxidation power of O$_{2}$
at temperatures above 400~\textdegree C.

\noindent Thus, the combination of a high temperature and a low flux
(large surface diffusion length), together with a high oxidation power
(f. e. NO$_{2}$ or plasma-activated O$_{2}$) is required to growth
smooth NiO layers.

\noindent 
\begin{table}
\noindent \centering{}\caption{Measured RMS roughness and grain size of S1~(1~sccm) and S2~(0.3~sccm).
No clear correlation is visible between the growth conditions and
the RMS roughness (R$_{RMS}$). The grain size is in general highest
for the 700~\textdegree C samples and in total higher for S2. \label{tab:RMS+GrainSize}}
\begin{tabular*}{8cm}{@{\extracolsep{\fill}}>{\centering}m{1cm}>{\centering}m{1.8cm}>{\centering}m{0.7cm}>{\centering}m{0.9cm}>{\centering}m{0.9cm}>{\centering}m{0.9cm}>{\centering}m{0.9cm}}
\toprule 
 & \textbf{T$_{g}${[}\textdegree C{]}} &  & \textbf{250} & \textbf{450} & \textbf{700} & \textbf{900}\tabularnewline
\cmidrule{2-7} 
\textbf{S1} & \textbf{R$_{RMS}${[}nm{]}} &  & 0.7 & 0.1 & 0.3 & 3.7\tabularnewline
\cmidrule{2-7} 
 & \textbf{size {[}nm{]}} &  & 30 & 15 & 50 & 67\tabularnewline
\midrule 
 & \textbf{T$_{g}${[}\textdegree C{]}} & \textbf{20} & \textbf{200} & \textbf{400} & \textbf{700} & \tabularnewline
\cmidrule{2-7} 
\textbf{S2} & \textbf{R$_{RMS}${[}nm{]}} & 0.7 & 0.25 & 1.6 & 0.4 & \tabularnewline
\cmidrule{2-7} 
 & \textbf{size {[}nm{]}} & 40 & 40 & 20 & 80 & \tabularnewline
\bottomrule
\end{tabular*}
\end{table}

\begin{figure}
\noindent \begin{centering}
\includegraphics[width=7.5cm]{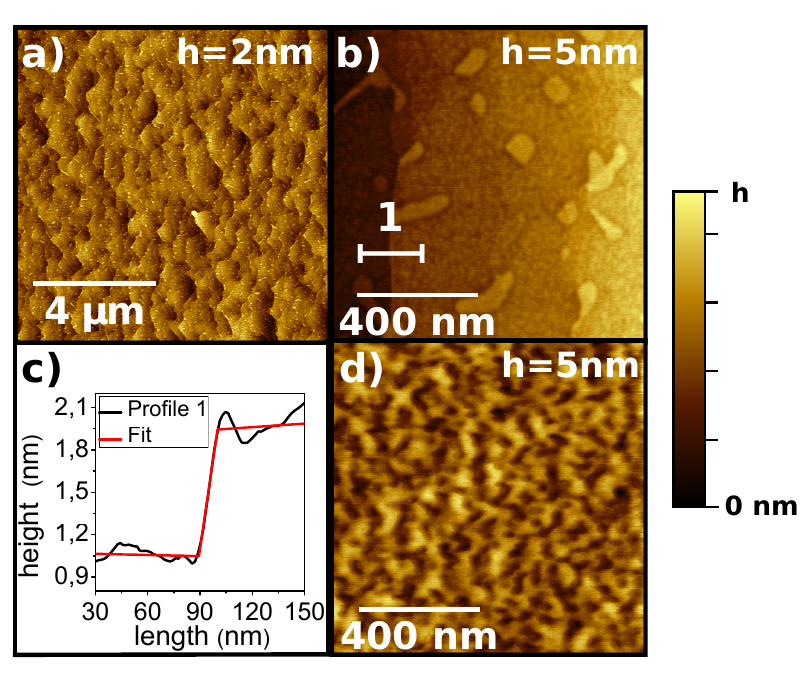}
\par\end{centering}
\caption{The AFM image of an annealed MgO(100) substrate (a), the magnified
image of a step (b) and the image of the grown layer NiO on it (d).
The line scan from (b) including a step fit is shown in (c). The substrate
steps are not adopted by the grown NiO layer.\label{fig:AFM_Anneal}}
\end{figure}

\noindent To further improve the surface morphology, different annealings
were tested to create a stepped layer surface. On the one hand, annealing
after growth of the NiO layer and, on the other hand, annealing before
growth of the MgO surface was investigated. Annealing in a tube furnace
after growth as suggested by Ohta et al.~\cite{Ohta} led to the
disappearance of the NiO peaks in all XRD measurements (not shown
here). We used an annealing temperature of 1200~\textdegree C for
30~min in air. \\
\\
The influence of annealing on the surface morphology of MgO has already
been investigated by Ahmed et al., who found terraces after 3 hours
at 1200~\textdegree C in an oxygen atmosphere.\cite{Ahmed_MgOAnnealing}
In our study, a stepped surface (see Fig.~\ref{fig:AFM_Anneal} a))
was reached after annealing the substrate four hours at 1150~\textdegree C
in a tube furnace with oxygen. Fig~\ref{fig:AFM_Anneal}~b) is a
magnified image of a) and shows the around 1~\textmu m wide steps.
On the steps small islands formed possibly due to a small diffusion
length. The step height is around 0.9~nm (see profile line in Fig.~\ref{fig:AFM_Anneal}~c))
which is similar to the expected height of 0.84~nm for a MgO double
step. After the growth of NiO (700~\textdegree C, 0.3~sccm) on this
substrate, no steps are visible (see Fig.~\ref{fig:AFM_Anneal} d)).
Compared to Fig.~\ref{fig:AFM_MgO100}~b) the morphology of the
layer changed according to the formation of interconnected islands
instead of grains. However, the steps are not adopted as shown in
Fig.~\ref{fig:AFM_Anneal} d). The roughness for the NiO layer on
the annealed MgO is 0.4~nm, which is similar to the sample roughness
of S1-700 and S2-700 (see Tab.~\ref{tab:RMS+GrainSize}). No difference
was seen in the XRD scan, either. Further improvement of the annealing
conditions for more defined steps without islands could lead to a
flat surface morphology without grains and improve the NiO layer.
This could be achieved by off\nobreakdash-cut substrates with a terrace
width for example of less than 80~nm for the S2-700 growth corresponding
to an off-cut angle above 0.3\textdegree .

\subsection{Raman quality metrics of NiO(100)/MgO(100)}

\noindent For rock-salt structures Raman spectroscopy is able to provide
information about the crystal quality as the first-order scattering
by optical phonons is forbidden due to the symmetry selection rules.\cite{Dietz_Raman,Mironova_Raman}
Crystal imperfections or the distortion below the N\'{e}el temperature
(T$_{N}$) could lead to the occurence of a first-order optical phonon
line (1P) in Raman spectra. However, Dietz et al.\cite{Dietz_Raman}
showed no significant increase of the 1P peak for temperatures below
T$_{N}$. Thus, the intensity of 1P Raman peak still correlates with
the layer quality. Since second-order Raman scattering by optical
phonons (2P) is allowed also for perfect crystals, the intensity of
the first-order Raman peak (I$_{1P}$) normalized to that of the second-order
peak (I$_{2P}$) can be utilized as an inverse figure of merit for
the crystal quality ($Q$): 
\begin{equation}
Q=\frac{I_{1P}}{I_{2P}}\label{eq:Q}
\end{equation}

\noindent where I$_{1P}$ and I$_{2P}$ are the integrated intensities
of the corresponding Raman peaks. Since I$_{1P}$ = 0 ($Q$ = 0) for
a perfect crystal, the value of $Q$ increases with the density of
crystal defects.
\begin{figure}
\noindent \begin{centering}
\includegraphics[width=8.5cm]{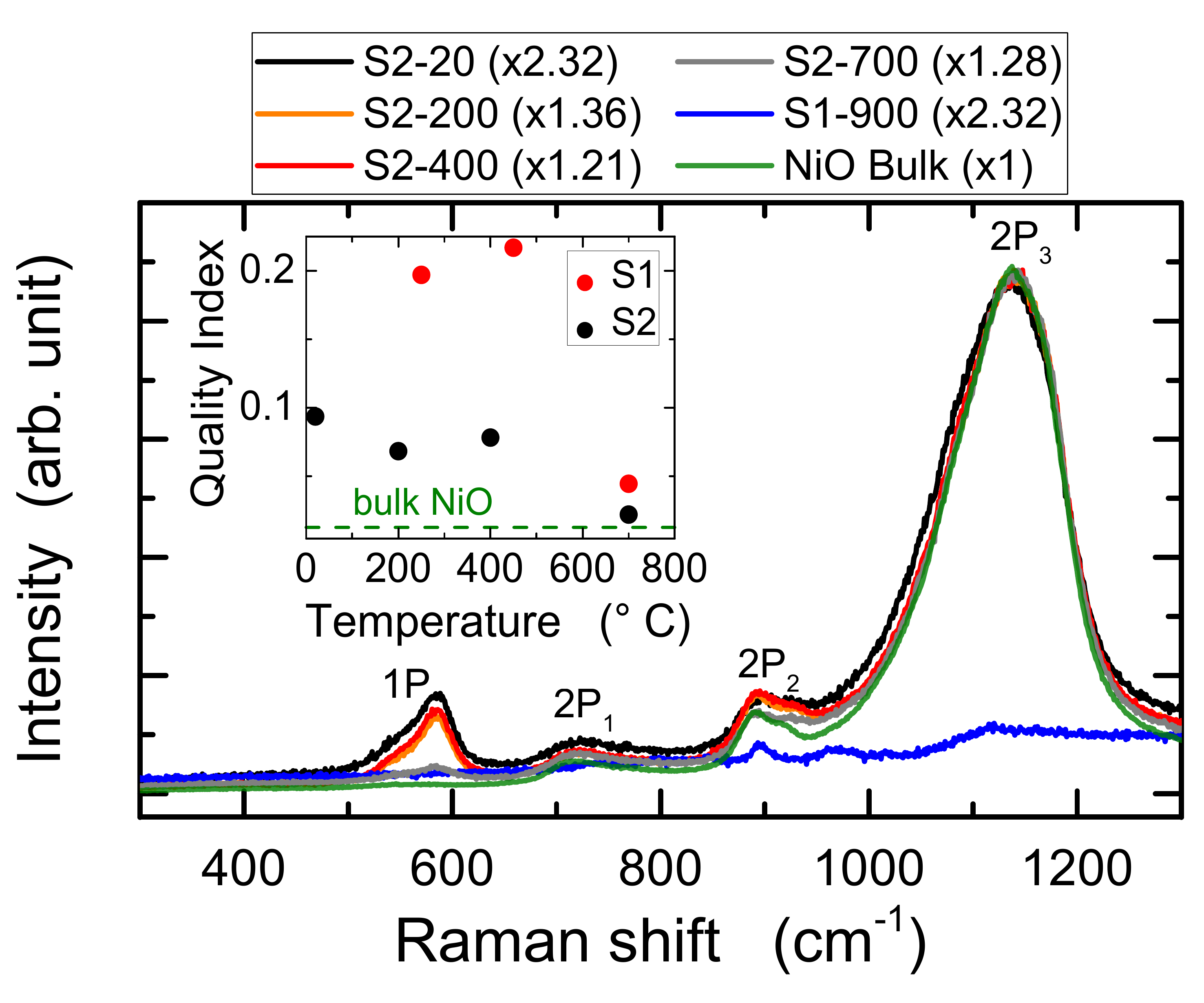}
\par\end{centering}
\caption{Raman spectra of sample series 2 and a bulk NiO reference sample.
The spectra are normalized to the intensity of one second\protect\nobreakdash-order
phonon scattering (2P$_{3}$) and reveal the correlation between the
growth temperature and the intensity of first\protect\nobreakdash-order
phonon scattering (1P). The applied scaling factor for each scan can
be found in the legend. The alloy (S1-900) shows no defined 2P$_{3}$
Raman peak. The inset shows the calculated quality index Q from samples
grown on MgO(100) up to 700~\textdegree C as a function of growth
temperature. The dashed line represents the quality index of a bulk
NiO reference sample. \label{fig:Raman_S2}}
\end{figure}

\begin{figure}
\noindent \begin{centering}
\includegraphics[width=8.5cm]{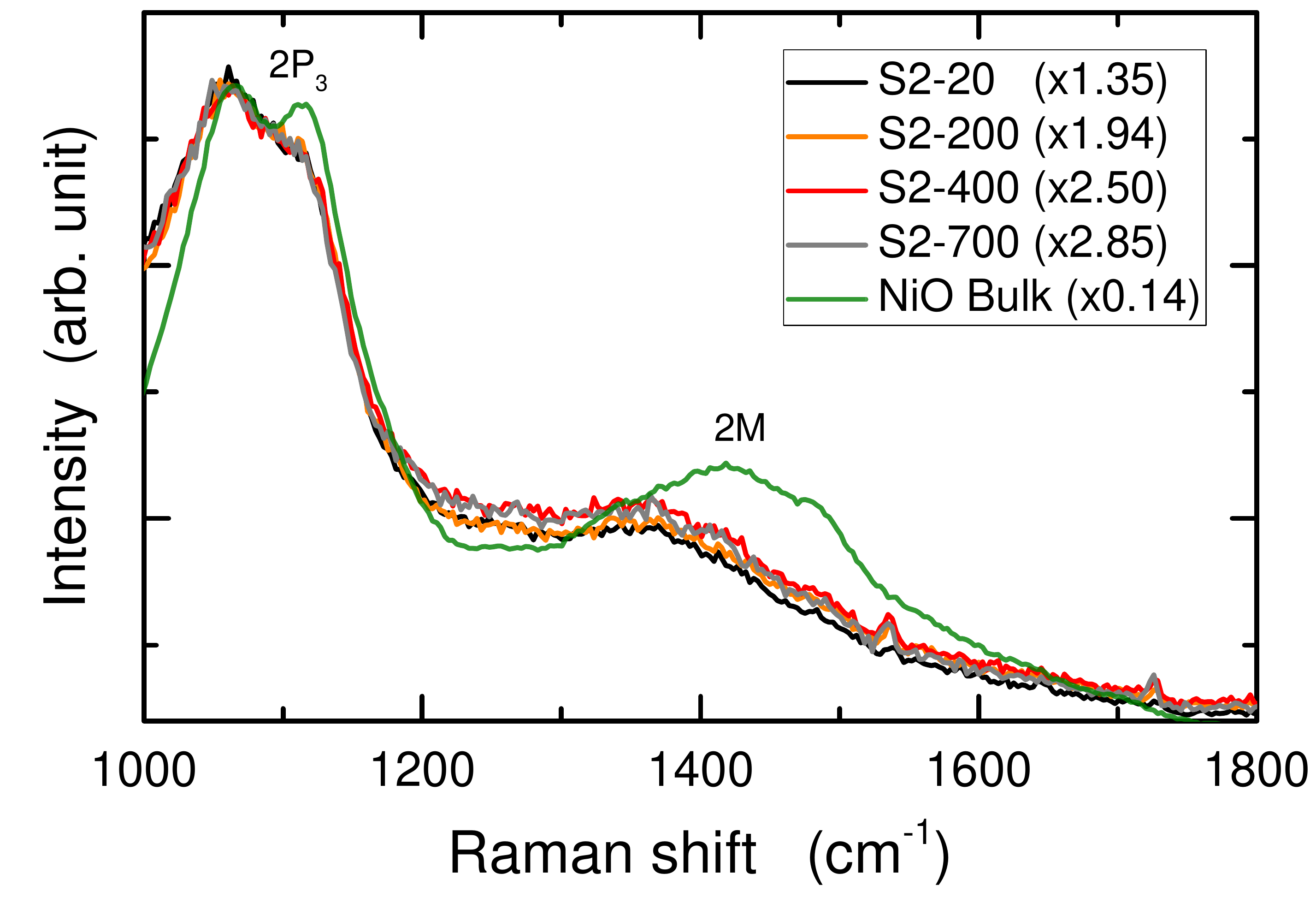}
\par\end{centering}
\caption{Raman spectra of all samples from series 2 measured at 100~\textdegree C
and excited at a wavelength of 405~nm. The shape of the 2P$_{3}$
peak in this Raman spectra is somewhat distorted by the spectral characteristic
of the particular Raman notch filter used for this excitation. The
spectra are normalized to the intensity of the 2P$_{3}$ peak and
exhibit 2M peaks with similar frequencies and relative intensities.
The spectrum of bulk NiO is shown for comparison. The applied scaling
factor for each scan can be found in the legend.\label{fig:Raman-2M}}
\end{figure}

\noindent For NiO three 2P peaks are visible in figure~\ref{fig:Raman_S2}.
The corresponding modes have been identified as the 2TO modes (2P$_{1}$),
the TO+LO modes (2P$_{2}$), and the 2LO modes (2P$_{3}$) by Mironova-Ulmane
et al.\cite{Mironova_Raman} For our calculation of $Q$ we used I$_{2P}$
as the integrated intensity of the 2P$_{3}$ peak. Raman spectra of
sample series 2 are shown in Fig.~\ref{fig:Raman_S2} with the intensity
normalized to the intensity of 2P$_{3}$. The decrease of the first-order
Raman peak (1P) clearly reveals the improvement of the NiO crystal
quality with increasing growth temperature. The NiO-MgO alloy (S1-900)
shows no defined Raman peak. The crystal quality index $Q$ defined
in Eq.~\ref{eq:Q} is shown in the inset in Fig.~\ref{fig:Raman_S2}
for all NiO films grown on MgO(100) as a function of growth temperature.
For both sample series (S1 and S2), $Q$ approaches at high temperature
the value of the nearly perfect bulk NiO reference sample (shown as
dashed line in the inset in Fig.~\ref{fig:Raman_S2}). The observed
improvement of the crystal quality with increasing growth temperature
is a frequently observed phenomenon for MBE grown films. The reason
for this improvement is given by enhanced diffusion length of adatoms
on the growth surface which increases the probability for the occurence
of step flow growth and the the formation of a thermodynamically stable
structure.\cite{Arthur_MBE} Consequently, the density of crystal
defects becomes reduced at elevated growth temperatures, especially
when the lattice mismatch between film and substrate is small. The
generally smaller $Q$ values (better crystal quality) observed for
sample series two (see inset in Fig.~\ref{fig:Raman_S2}) can be
explained by the smaller oxygen flux during growth. The lower surface
coverage by oxygen atoms also enables a higher surface mobility of
adatoms which is beneficial for the obtained crystal quality (as explained
above). Furthermore, the higher crystal quality of NiO films prepared
with smaller oxygen flux indicates that oxygen vacancies are not the
main source of the crystal imperfections in the investigated NiO films.
Our results demonstrate that Raman spectroscopy can be utilized in
a very efficient manner to evaluate the crystal quality of NiO films.
A similar approach has already been used to estimate the amount of
crystal defects in SrTiO3 films.\cite{Tenne_SrTiO3} In contrast to
the XRD measurements, Raman spectroscopy is sensitive to imperfections
of the crystal structure in the investigated NiO films since the Raman
selection rules are influenced by the local crystal symmetry rather
than by the flatness of the substrate. The local symmetry, however,
is only influenced by lattice distortions and defects or grain boundaries.\\
\\
Regarding the magnetic ordering, we investigated second\nobreakdash-order
magnon scattering using a 405\nobreakdash-nm diode laser for excitation
since the corresponding two magnon (2M) peak could not be observed
for excitation at 325~nm.\cite{Aytan_2MSuppression} The 2M peak
is in general observable below the N\'{e}el temperature as a signature
of antiferromagnetic ordering. Based on a detailed investigation of
the dispersion for different directions in NiO, Betto et al.\cite{Betto_J}
revealed a reduction of the leading superexchange parameter for films
compared to bulk NiO, presumably due to strain. For our films, independent
of the growth temperature, a 2M peak is measurable as a fingerprint
of antiferromagnetic ordering (see Fig.~\ref{fig:Raman-2M}). The
observed redshift with respect to bulk NiO reflects the expected strain\nobreakdash-induced
decrease in the superexchange parameter,\cite{Betto_J} which is proportional
to the 2M frequency.\cite{Gandhi_Jvs2M}

\subsection{Optical properties}

\noindent The optical properties are investigated by ellipsometry
measurements of the NiO films on MgO(100) substrates. For the evaluation
of the data, a three-layer model was used: Two NiO layers on a MgO
substrate. The topmost NiO layer is needed to describe the surface
roughness of the NiO and the other one describes the actual NiO film
properties. The roughness layer was modeled by the Bruggeman effective
medium approach assuming 50~\% voids. It has a thickness of 2\nobreakdash-7~nm.
In most of the actual NiO layers voids were needed as well for the
description of the optical properties, however only to a volume fraction
below 7~\%. There is a trend visible that less voids have to be assumed
for higher growth temperatures, which indicates a better crystal quality.
For the S2 samples an introduction of a certain inhomogeneity, that
could be related to thickness fluctuation on the sample, was found
to produce a better fit than a model solely based on voids and surface
roughness. However, the model properties inhomogeneity, voids, and
surface roughness are intertwined inextricably.\cite{Elli1,Elli2}
The relative differences of surface roughness as found by AFM~(Tab.~\ref{tab:RMS+GrainSize})
presently is not reflected by the roughness layer thickness d\textbf{$_{R}$}
(Tab.~\ref{tab:elli Data}), determined by ellipsometry measurements.
If however, the inhomogeneity would be removed from the model, the
value for $d_{R}$ would follow more closely the trends found by AFM
at the expense of a slightly decreased quality of the overall agreement
between optical model and experimental data. All results are tabulated
in table~\ref{tab:elli Data}.
\begin{table}
\caption{The modeled data of the ellipsometry measurements for all layers grown
on MgO(100). The thickness is divided in the layer to describe the
surface roughness (d$_{R}$) with 50~\% voids and the remaining NiO
layer (d$_{NiO}$). Furthermore, the data include the modeled voids
for the remaining NiO layer of each sample and the inhomogeneity (I)
for the second series. \label{tab:elli Data}}
\noindent \centering{}%
\begin{tabular*}{8cm}{@{\extracolsep{\fill}}>{\centering}m{1.3cm}>{\raggedright}m{2cm}>{\centering}m{0.9cm}>{\centering}m{0.9cm}>{\centering}m{0.9cm}>{\centering}m{0.9cm}}
\toprule 
 & \textbf{T$_{g}${[}\textdegree C{]}} &  & \textbf{250} & \textbf{450} & \textbf{700}\tabularnewline
\cmidrule{2-6} 
\textbf{S1} & \textbf{d$_{R}$ {[}nm{]}} &  & 7.3 & 4.5 & 7.0\tabularnewline
\cmidrule{2-6} 
 & \textbf{d$_{NiO}$ {[}nm{]}} &  & 47.3 & 23.1 & 41.4\tabularnewline
\cmidrule{2-6} 
 & \textbf{voids {[}\%{]}} &  & 7.2 & 0.0 & 3.4\tabularnewline
\midrule 
 & \textbf{T$_{g}${[}\textdegree C{]}} & \textbf{20} & \textbf{200} & \textbf{400} & \textbf{700}\tabularnewline
\cmidrule{2-6} 
 & \textbf{d$_{R}$ {[}nm{]}} & 4.0 & 4.0 & 7.0 & 2.0\tabularnewline
\cmidrule{2-6} 
\textbf{S2} & \textbf{d$_{NiO}$ {[}nm{]}} & 61.2 & 60.4 & 53.1 & 60.7\tabularnewline
\cmidrule{2-6} 
 & \textbf{voids {[}\%{]}} & 7.3 & 4.5 & 4.6 & 3.7\tabularnewline
\cmidrule{2-6} 
 & \textbf{I {[}\%{]}} & 7 & 17 & 31 & 13\tabularnewline
\bottomrule
\end{tabular*}
\end{table}

\begin{figure}
\noindent \begin{centering}
\includegraphics[width=7.5cm]{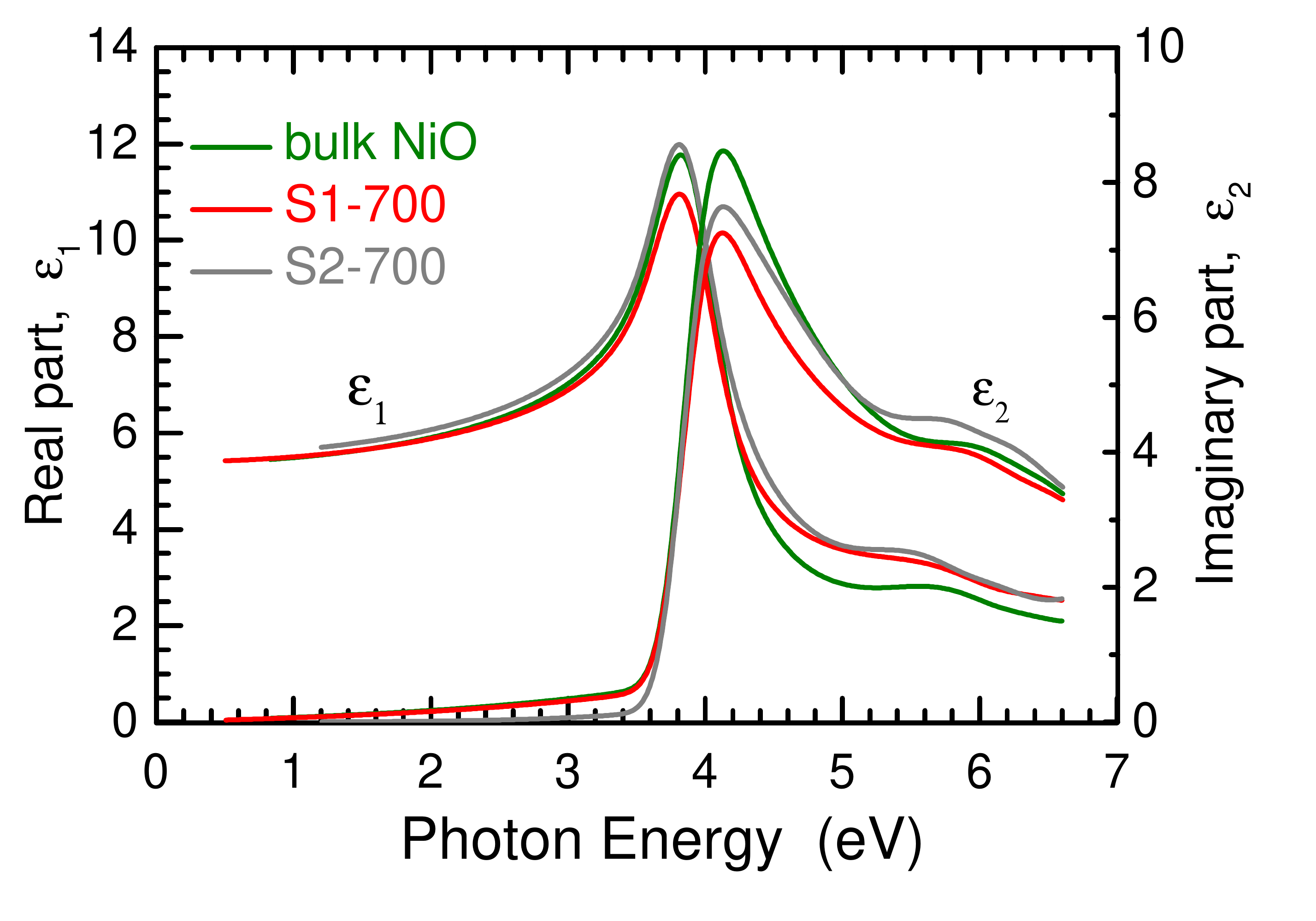}
\par\end{centering}
\caption{Dielectric functions of the bulk NiO and both thin film samples grown
at 700\textdegree C. Comparing an oxygen poor sample (S2-700) with
a stoichiometric one (S1-700). \label{fig:diel Function}}
\end{figure}

\noindent The combined modeled thickness of the two NiO layers is
similar to the thicknesses measured by XRR. Furthermore, all layers
showed an onset of strong interband absorption around 3.6~eV, which
is in excellent agreement with earlier experimental findings.\cite{Opt_Kang,Opt_Roedl,Opt_Ghosh,Opt_Battiato}
All samples yield spectra with sharp and pronounced absorption features
in the imaginary parts of their DFs which is a hint towards high crystalline
and morphological quality. The DFs of the bulk NiO crystal and samples
S1-700 and S2-700 are displayed in Fig.~\ref{fig:diel Function}.
The increasing value of the real part of the DF for the oxygen poor
sample (S2-700) in the transparency region below about 3.5eV is in
agreement with Ref.~{[}\citen{Becker_NixO}{]}, however the general
line shape and energy positions of distinguished features is in very
good agreement to those of Refs.~{[}\citen{Opt_Kang,Opt_Roedl,Opt_Ghosh,Opt_Battiato}{]}.
The influence of different nickel oxide compositions (Ni$_{x}$O)
for example on the refractive index has been investigated by Becker
et al. for x ranging from 0.9 to 1.45.\cite{Becker_NixO}

\subsection{Electrical properties}

\begin{figure}
\includegraphics[width=5.5cm]{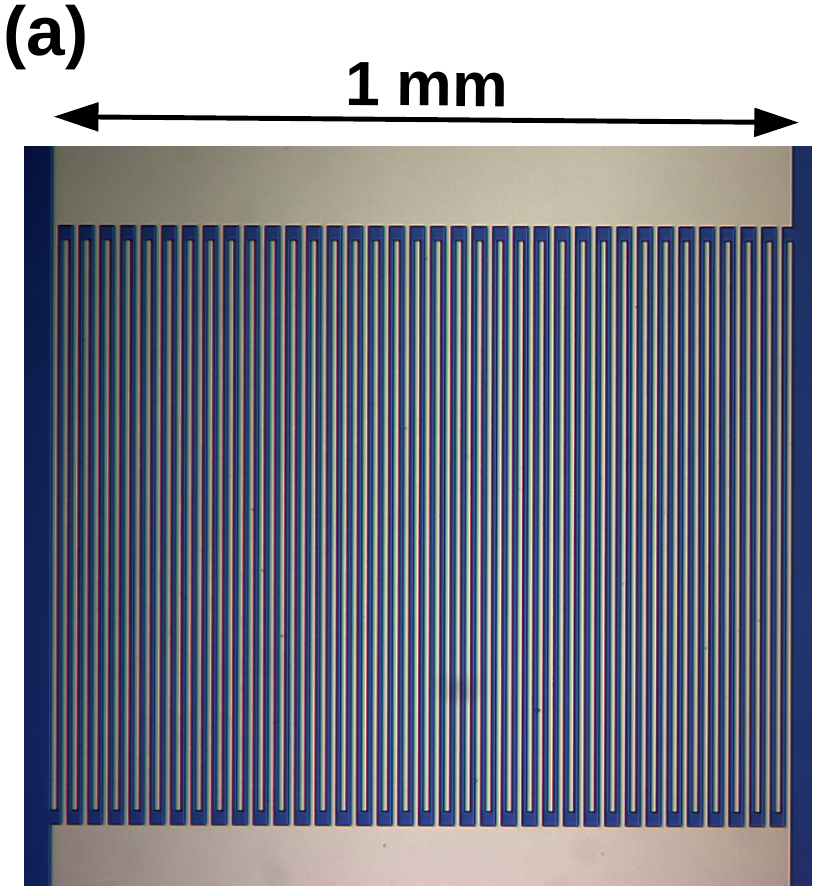}

\includegraphics[width=7cm]{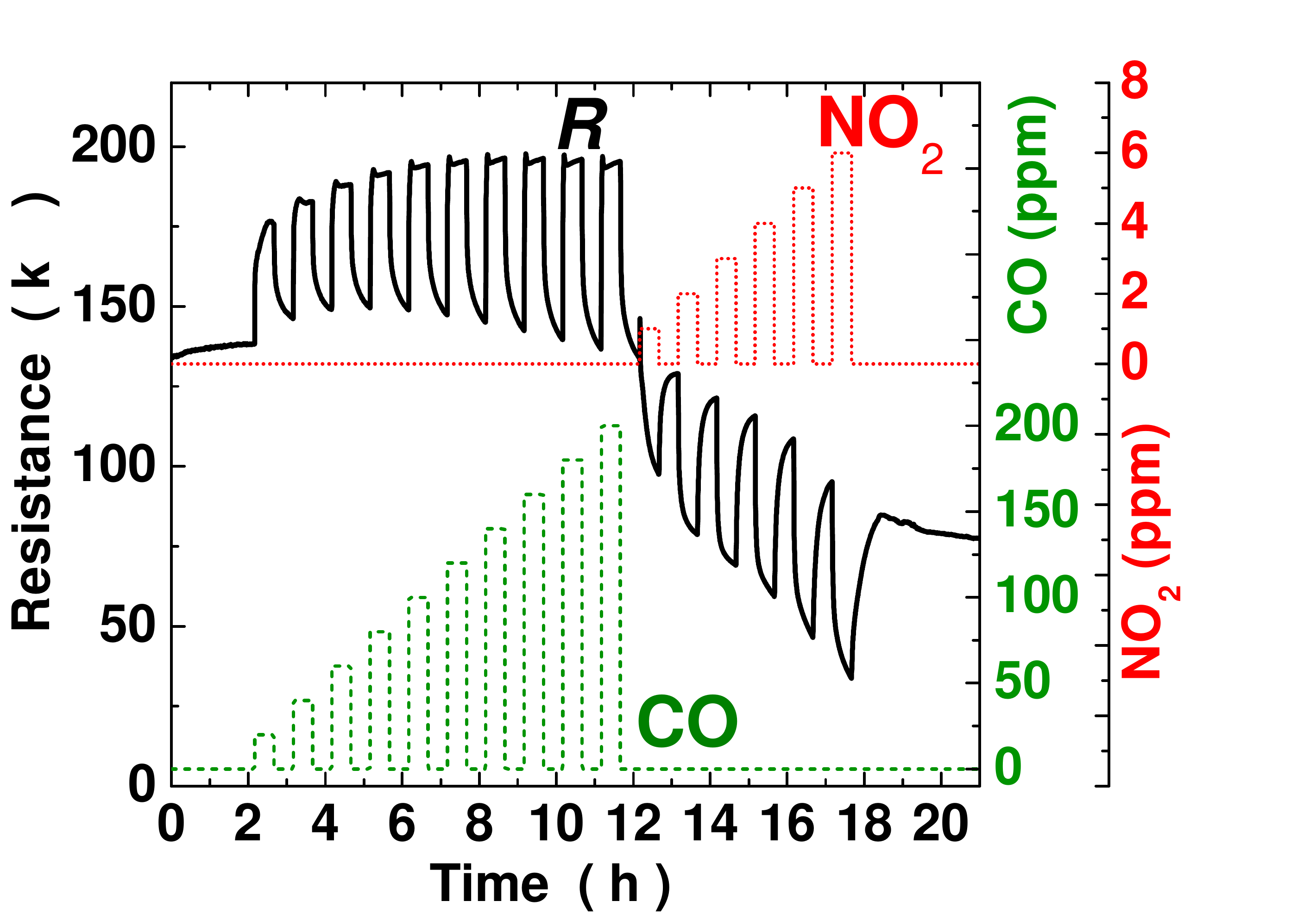}

\caption{Interdigitated contact structure (a) for the resistance measurement
shown in (b). The resistance is measured at a temperature of 220\textdegree C
as a function of a time dependent exposure to synthetic air with defined
concentration of the reducing gas CO (acting as surface donor) and
the oxidizing gas NO$_{2}$ (acting as surface acceptor). \label{fig:R}}
\end{figure}

\noindent The idcs used in this study, reduced the measured resistance
with respect to the sheet resistance by a defined geometry factor
of $1.3\times10^{-4}$ for the structures shown in Fig.~\ref{fig:R}~a)
and by a factor of $10^{-3}$ using the structures shown in Ref~{[}\citen{Contact_Patterns}{]}.
Based on the measured resistance of the idcs, the calculated sheet
resistances at room temperature for all samples grown on MgO(100)
are higher than $10^{9}\,\varOmega$, indicating no significant bulk
conductivity. The low bulk conductivity may be due to a too low concentration
of unintentional acceptors (e.g. Ni vacancies) or the existence of
a high concentration of compensating unintentional donors. Due to
the high resistance of all samples any attempt to determine the conductivity
type by Hall, capacitance\nobreakdash-voltage or Seebeck measurements,
failed.

\noindent As an alternative for determining the conductivity type
we measured the resistance change of the idcs upon exposure to reducing
or oxidizing gases at elevated temperatures \textemdash{} a behavior
that is exploited in oxide-semiconductor based gas sensors. Oxidizing
gases, such as NO$_{2}$, act as surface acceptors, whereas reducing
gases, such as CO, act as surface donors.\cite{Kim2014} Consequently,
exposure to oxidizing gases would reduce the number of electrons in
$n$-type oxides and thus increase their resistance, whereas it would
increase the number of holes and hence decrease the resistance of
$p$-type oxides.\cite{Barsan_gasssensing,Kim2014} The effect of
reducing gases is opposite to that of oxidizing gases by either acting
directly as surface donors or by reacting with and thus reducing the
number of adsorbed surface-accepting oxygen.\cite{Barsan_gasssensing}
Fig.~\ref{fig:R} (b) shows the resistance of the idcs (a) for S2\nobreakdash-400
as a function of NO$_{2}$ and CO content in synthetic air without
humidity. The increasing resistance under CO exposure and decreasing
resistance under NO$_{2}$ exposure consistently demonstrates $p$-type
conductivity of our NiO sample.

\section{Summary and conclusions}

\noindent We investigated the growth of NiO layers by plasma-assisted
MBE on MgO(100), MgO(110) and MgO(111). On MgO(100) and MgO(110) the
NiO layers were single crystalline with (100) and (110) orientation,
respectively, having an epitaxial cube\nobreakdash-on\nobreakdash-cube
relationship to the substrate. The NiO layer on MgO(111), in contrast,
consisted of (111) and (001) oriented domains, likely related to the
rough substrate surface. The influence of growth conditions on film
properties was investigated on MgO(100) by the variation of growth
temperature and active oxygen flux: Single crystalline growth was
maintained on MgO(100) in the entire range of investigated growth
temperatures, i.e. from 20~\textdegree C to 900~\textdegree C, and
different oxygen fluxes (including molecular oxygen), that all resulted
in O-rich and Ni-limited growth conditions. Layers grown under vastly
different conditions were fairly stoichiometric with an oxygen content
in the range of 47~at.\% to 50~at.\%. All layers were transparent
to the visible light and had a RMS surface roughness typically below
1~nm. Only the growth at 900\textdegree ~C or with molecular oxygen
led to significantly higher roughnesses. At the growth temperature
of 900~\textdegree C Mg diffused from the substrate into the layer,
resulting in a Mg$_{x}$Ni$_{1-x}$O(100) film.

\noindent The intensity ratios of quasi-forbidden one-phonon to allowed
two-phonon Raman peaks was introduced as a new quality metrics, which
indicated the highest layer quality for a high growth temperature
of 700~\textdegree C and a low oxygen flux. 

\noindent Therefore, for optimum crystal quality a growth of NiO on
MgO(100) at a temperature $\ge700$~\textdegree C but $<900$~\textdegree C
using a low, plasma-activated oxygen flux is recommended by our data.
A comparison to available literature suggests the importance of a
low growth rate. A further improvement in crystal quality by using
higher growth temperatures may be enabled by a thermodynamically more
stable substrate material, such as sapphire (Al$_{2}$O$_{3}$), that
does not diffuse into the growing film. The epitaxy by ion-beam sputtering
on a- and c-plane Al$_{2}$O$_{3}$ resulted in NiO(111) films with
two rotational domains, whereas that on m-plane Al$_{2}$O$_{3}$
resulted in NiO(110) films consisting of multiple tilt domains.\cite{Becker2017}
Preparation of atomically bistepped (step heigth 0.433~nm) c-plane
Al$_{2}$O$_{3}$ surfaces by annealing prior to growth has been shown
to prevent the formation of rotational domains, enabling the growth
of single crystalline NiO(111) films thereon as demonstrated by pulsed
laser deposition.\cite{NiO_Al2O3}

\noindent AFM images of our films showed a surface composed of grains
with larger size for higher temperatures or lower activated oxygen
flux, indicating an increasing surface diffusion length. Together
with the Raman-derived quality metrics a correlation between a high
layer quality and a high surface diffusion length is found.

\noindent All grown layers were semi-insulating with sheet resistance
$>1$~G$\Omega$, indicating a too low concentration of unintentional
acceptors (e.g. Ni vacancies) or the existence of a sufficiently high
concentration of compensating unintentional donors. Nevertheless,
\textit{p}\nobreakdash-type conductivity was confirmed at elevated
temperature by gas response measurements of a representative layer,
demonstrating the suitability of single crystalline NiO thin films
for fundamental gas sensor research. 
\begin{acknowledgments}
\noindent We would like to thank S. Rauwerdink and W. Seidel for cleanroom
processing, H.-P. Sch\"onherr for MBE support, U. Jahn for measuring
and evaluating the EDX data, and J.~M.~J.~Lopes for critically
reading the manuscript.. This work was performed in the framework
of GraFOx, a Leibniz-ScienceCampus partially funded by the Leibniz
association. M.~B., P.~F., and J.~F. gratefully acknowledges financial
support by the Leibniz Association.
\end{acknowledgments}

\bibliographystyle{unsrtnat}
\bibliography{MgO_NiO_Paper}

\begin{thebibliography}{51}
\providecommand{\natexlab}[1]{#1}
\providecommand{\url}[1]{\texttt{#1}}
\expandafter\ifx\csname urlstyle\endcsname\relax
  \providecommand{\doi}[1]{doi: #1}\else
  \providecommand{\doi}{doi: \begingroup \urlstyle{rm}\Url}\fi
  \providecommand{\nmbibLink}[3]{#3}

\bibitem[Ohta et~al.(2003)Ohta, Kamiya, Kamiya, Hirano, and Hosono]{Ohta}
\nmbibLink{Ohta}{authors}{H.~Ohta, M.~Kamiya, T.~Kamiya, M.~Hirano, and
  H.~Hosono}, \emph{Thin Solid Films}, {\bf 445}  (2003).
\newblock \doi{10.1016/S0040-6090(03)01178-7}.

\bibitem[Rao and Smakula(1965)]{Rao_NiO-prop}
\nmbibLink{Rao_NiO-prop}{authors}{K.~V. Rao and A.~Smakula}, \emph{J. Appl.
  Phys.}, {\bf 36}\penalty0 (6),  2031  (1965).
\newblock \doi{10.1063/1.1714397}.

\bibitem[Zhang et~al.(2018)Zhang, Li, Hoye, MacManus-Driscoll, Budde,
  Bierwagen, Wang, Du, Wahila, Piper, Lee, Edwards, Dhanak, and
  Zhang]{Zhang2018}
\nmbibLink{Zhang2018}{authors}{J.-Y. Zhang, W.~Li, R.~L.~Z. Hoye,
  J.~MacManus-Driscoll, M.~Budde, O.~Bierwagen, L.~Wang, Y.~Du, M.~Wahila,
  L.~F.~J. Piper, T.-L. Lee, H.~Edwards, V.~R. Dhanak, and H.~Zhang}, \emph{J.
  Mater. Chem. C}, {\bf 6},  2275--2282  (2018).
\newblock \doi{10.1039/C7TC05331B}.

\bibitem[Schuler et~al.(2005)Schuler, Ederer, and Itza-Oritz]{Schuler}
\nmbibLink{Schuler}{authors}{T.~M. Schuler, D.~L. Ederer, and S.~Itza-Oritz},
  \emph{Phys. Rev. B}, {\bf 71}\penalty0 (115113),  1  (2005).
\newblock \doi{10.1103/PhysRevB.71.115113}.

\bibitem[Massidda et~al.(1999)Massidda, Posternak, Baldereschi, and
  Resta]{Massidda_Distortion}
\nmbibLink{Massidda_Distortion}{authors}{S.~Massidda, M.~Posternak,
  A.~Baldereschi, and R.~Resta}, \emph{Phys. Rev. Lett.}, {\bf 82}\penalty0
  (2),  430  (1999).
\newblock \doi{10.1103/PhysRevLett.82.430}.

\bibitem[Hoshiya et~al.(1994)Hoshiya, Komuro, Mitsuoka, and Sugita]{Hoshiya}
\nmbibLink{Hoshiya}{authors}{H.~Hoshiya, M.~Komuro, K.~Mitsuoka, and
  Y.~Sugita}, \emph{IEEE Transl. J. Magn. Jpn.}, {\bf 9}\penalty0 (6),  236
  (1994).
\newblock \doi{10.1109/TJMJ.1994.4565986}.

\bibitem[Becker et~al.(2017{\natexlab{a}})Becker, Polity, and Klar]{Becker2017}
\nmbibLink{Becker2017}{authors}{M.~Becker, A.~Polity, and P.~J. Klar},
  \emph{Journal of Applied Physics}, {\bf 122}\penalty0 (17),  175303  (2017).
\newblock \doi{10.1063/1.4991601}.
\newblock URL \url{https://doi.org/10.1063/1.4991601}.

\bibitem[Kim and Lee(2014)]{Kim2014}
\nmbibLink{Kim2014}{authors}{H.-J. Kim and J.-H. Lee}, \emph{Sens. Actuators,
  B}, {\bf 192},  607  (2014).
\newblock \doi{http://dx.doi.org/10.1016/j.snb.2013.11.005}.

\bibitem[Miller et~al.(2014)Miller, Akbar, and Morris]{GasSen_Overwiev}
\nmbibLink{GasSen_Overwiev}{authors}{D.~R. Miller, S.~A. Akbar, and P.~A.
  Morris}, \emph{Sens. Actuators B}, {\bf 204},  250  (2014).
\newblock \doi{10.1016/j.snb.2014.07.074}.

\bibitem[Zhang et~al.(2016)Zhang, Xi, Blamire, and Egdell]{Zhang_pType}
\nmbibLink{Zhang_pType}{authors}{K.~H.~L. Zhang, K.~Xi, M.~G. Blamire, and
  R.~G. Egdell}, \emph{J. Phys. Condens. Matter}, {\bf 28}\penalty0 (38),  1
  (2016).
\newblock \doi{10.1088/0953-8984/28/38/383002}.

\bibitem[von Wenckstern et~al.(2015)von Wenckstern, Splith, Lanzinger, Schmidt,
  Müller, Schlupp, Karsthof, and Grundmann]{Wenckstern2015}
\nmbibLink{Wenckstern2015}{authors}{H.~von Wenckstern, D.~Splith, S.~Lanzinger,
  F.~Schmidt, S.~Müller, P.~Schlupp, R.~Karsthof, and M.~Grundmann},
  \emph{Adv. Electron. Mater.}, {\bf 1},  1400026  (2015).
\newblock \doi{10.1002/aelm.201400026}.

\bibitem[Irwin et~al.(2007)Irwin, Buchholz, Hains, Chang, and Marks]{Irwin}
\nmbibLink{Irwin}{authors}{M.~D. Irwin, D.~B. Buchholz, A.~W. Hains, R.~P.~H.
  Chang, and T.~J. Marks}, \emph{Proc. Natl. Acad. Sci. U.S.A}, {\bf
  105}\penalty0 (8),  2783  (2007).
\newblock \doi{10.1073/pnas.0711990105}.

\bibitem[Warot et~al.(2001{\natexlab{a}})Warot, Snoeck, Baul\`{e}s, Ousset,
  Casanove, Dubourg, and Bobo]{Warot_Growth_MgO001_MgO110}
\nmbibLink{Warot_Growth_MgO001_MgO110}{authors}{B.~Warot, E.~Snoeck,
  P.~Baul\`{e}s, J.~C. Ousset, M.~J. Casanove, S.~Dubourg, and J.~F. Bobo},
  \emph{Appl. Surf. Sci.}, {\bf 177}\penalty0 (4),  287  (2001).
\newblock \doi{10.1016/S0169-4332(01)00223-9}.

\bibitem[Becker et~al.(2017{\natexlab{b}})Becker, Polity, and
  Klar]{Becker_NixO}
\nmbibLink{Becker_NixO}{authors}{M.~Becker, F.~Michel~A. Polity, and P.~J.
  Klar}, \emph{Phys. Status Solidi B}, {\bf 255}\penalty0 (1700463)  (2017).
\newblock \doi{10.1002/pssb.201700463}.

\bibitem[Murphy and Hutchins(1995)]{NiO_Ox_states}
\nmbibLink{NiO_Ox_states}{authors}{T.~P. Murphy and M.~G. Hutchins}, \emph{Sol.
  Energy Mater. Sol. Cells}, {\bf 39}\penalty0 (2-4),  377  (1995).
\newblock \doi{10.1016/0927-0248(96)80003-1}.

\bibitem[Peacor and Hibma(1994)]{Peacor}
\nmbibLink{Peacor}{authors}{S.~D. Peacor and T.~Hibma}, \emph{Surf. Sci.}, {\bf
  301},  11  (1994).
\newblock \doi{10.1016/0039-6028(94)91283-1}.

\bibitem[Tachiki et~al.(2000)Tachiki, Hosomi, and Kobayashi]{Tachiki}
\nmbibLink{Tachiki}{authors}{M.~Tachiki, T.~Hosomi, and T.~Kobayashi},
  \emph{Jpn. J. Appl. Phys.}, {\bf 39 Part 1}\penalty0 (4A),  1817  (2000).
\newblock \doi{10.1143/JJAP.39.1817}.

\bibitem[Manders et~al.(2013)Manders, Tsang, Hartel, Lai, Chen, Amb, Reynolds,
  and So]{Manders_sNiO}
\nmbibLink{Manders_sNiO}{authors}{J.~R. Manders, S.-W. Tsang, M.~J. Hartel,
  T.-H. Lai, S.~Chen, C.~M. Amb, J.~R. Reynolds, and F.~So}, \emph{Adv. Funct.
  Mater}, {\bf 23}\penalty0 (23),  2993  (2013).
\newblock \doi{10.1002/adfm.201202269}.

\bibitem[Lind et~al.(1992)Lind, Berry, Chem, Mathias, and Testardi]{Lind}
\nmbibLink{Lind}{authors}{D.~M. Lind, S.~D. Berry, G.~Chem, H.~Mathias, and
  L.~R. Testardi}, \emph{Phys. Rev. B}, {\bf 45}\penalty0 (4),  1838  (1992).
\newblock \doi{10.1103/PhysRevB.45.1838}.

\bibitem[Mares et~al.(2010)Mares, Boutwell, Wei, Scheurer, and
  Schoenfeld]{Mares_NiMgO}
\nmbibLink{Mares_NiMgO}{authors}{J.~W. Mares, R.~C. Boutwell, M.~Wei,
  A.~Scheurer, and W.~V. Schoenfeld}, \emph{Appl. Phys. Lett.}, {\bf 97},
  161113  (2010).
\newblock \doi{10.1063/1.3503634}.

\bibitem[Rombach et~al.(2016)Rombach, Papadogianni, Mischo, Cimalla, Kirste,
  Ambacher, Berthold, Krischok, Himmerlich, Selve, and Bierwagen]{Rombach2016}
\nmbibLink{Rombach2016}{authors}{J.~Rombach, A.~Papadogianni, M.~Mischo,
  V.~Cimalla, L.~Kirste, O.~Ambacher, T.~Berthold, S.~Krischok, M.~Himmerlich,
  S.~Selve, and O.~Bierwagen}, \emph{Sensors and Actuators B: Chemical}, {\bf
  236},  909  (2016).
\newblock ISSN 0925-4005.
\newblock \doi{10.1016/j.snb.2016.03.079}.

\bibitem[Wang et~al.(2011)Wang, Fang, Long, Li, Mo, Huang, Zhou, and
  Zhao]{Wang_NiO/MgO/GaN_Diode}
\nmbibLink{Wang_NiO/MgO/GaN_Diode}{authors}{H.~Wang, G.~Fang, H.~Long, S.~Li,
  X.~Mo, H.~Huang, H.~Zhou, and X.~Zhao}, \emph{Semicond. Sci. Technol.}, {\bf
  26},  125015  (2011).
\newblock \doi{10.1088/0268-1242/26/12/125015}.

\bibitem[Warot et~al.(2002{\natexlab{a}})Warot, Snoeck, Baul\`{e}s, Ousset,
  Casanove, Dubord, and Bobo]{Warot_Islands_MgO111}
\nmbibLink{Warot_Islands_MgO111}{authors}{B.~Warot, E.~Snoeck, P.~Baul\`{e}s,
  J.~C. Ousset, M.~J. Casanove, S.~Dubord, and J.~F. Bobo}, \emph{J. Cryst
  Growth.}, {\bf 234}\penalty0 (4),  704  (2002).
\newblock \doi{10.1016/S0022-0248(01)01767-5}.

\bibitem[Warot et~al.(2001{\natexlab{b}})Warot, Snoeck, Baul\`{e}s, Ousset,
  Casanove, Dubourg, and Bobo]{Warot_Morph_MgO110}
\nmbibLink{Warot_Morph_MgO110}{authors}{B.~Warot, E.~Snoeck, P.~Baul\`{e}s,
  J.~C. Ousset, M.~J. Casanove, S.~Dubourg, and J.~F. Bobo}, \emph{J. Cryst
  Growth.}, {\bf 224}\penalty0 (3-4),  309  (2001).
\newblock \doi{10.1016/S0022-0248(01)01017-X}.

\bibitem[Warot et~al.(2002{\natexlab{b}})Warot, Snoeck, Ousset, Casanove,
  Dubord, and Bobo]{Warot_Morpho_all_Orientations}
\nmbibLink{Warot_Morpho_all_Orientations}{authors}{B.~Warot, E.~Snoeck, J.~C.
  Ousset, M.~J. Casanove, S.~Dubord, and J.~F. Bobo}, \emph{Appl. Surf. Sci.},
  {\bf 188}\penalty0 (1-2),  151  (2002).
\newblock \doi{10.1016/S0169-4332(01)00725-5}.

\bibitem[Goldhahn(2003)]{DF_Goldhahn}
\nmbibLink{DF_Goldhahn}{authors}{R.~Goldhahn}, \emph{Acta Phys. Pol A}, {\bf
  104},  123  (2003).
\newblock \doi{10.12693/APhysPolA.104.123}.

\bibitem[Bruggeman(1935)]{Bruggeman}
\nmbibLink{Bruggeman}{authors}{D.~A.~G. Bruggeman}, \emph{Ann. Phys. (Berl.)},
  {\bf 416}\penalty0 (7),  636  (1935).
\newblock \doi{10.1002/andp.19354160705}.

\bibitem[Feneberg et~al.(2016)Feneberg, Nixdorf, Lidig, Goldhahn, Galazka,
  Bierwagen, and Speck]{Feneberg_In2O3}
\nmbibLink{Feneberg_In2O3}{authors}{M.~Feneberg, J.~Nixdorf, C.~Lidig,
  R.~Goldhahn, Z.~Galazka, O.~Bierwagen, and J.~S. Speck}, \emph{Phys. Rev. B},
  {\bf 93}\penalty0 (4),  045203  (2016).
\newblock \doi{10.1103/PhysRevB.93.045203}.

\bibitem[Barsan et~al.(2010)Barsan, Simion, Heine, Pokhrel, and
  Weimar]{Barsan_gasssensing}
\nmbibLink{Barsan_gasssensing}{authors}{N.~Barsan, C.~Simion, T.~Heine,
  S.~Pokhrel, and U.~Weimar}, \emph{J. Electroceramics}, {\bf 25}\penalty0 (1),
   11  (2010).
\newblock \doi{10.1007/s10832-009-9583-x}.

\bibitem[Fewster(1999)]{Fewster_Fringes}
\nmbibLink{Fewster_Fringes}{authors}{P.~F. Fewster}, \emph{Rep. Prog. Phys.},
  {\bf 59},  1339  (1999).
\newblock \doi{10.1088/0034-4885/59/11/001}.

\bibitem[James and Hibma(1999)]{James_Strain}
\nmbibLink{James_Strain}{authors}{M.~A. James and T.~Hibma}, \emph{Surf. Sci.},
  {\bf 433},  718  (1999).
\newblock \doi{10.1016/S0039-6028(99)00476-8}.

\bibitem[Boutwell et~al.(2012)Boutwell, Wei, Scheurer, Mares, and
  Schoenfeld]{Boutwell_NiMgO}
\nmbibLink{Boutwell_NiMgO}{authors}{R.~C. Boutwell, M.~Wei, A.~Scheurer, J.~W.
  Mares, and W.~V. Schoenfeld}, \emph{Thin Solid Films}, {\bf 520}\penalty0
  (13),  4302  (2012).
\newblock \doi{10.1016/j.tsf.2012.02.065}.

\bibitem[Moram and Vickers(2009)]{Vickers_X-ray_nitrides}
\nmbibLink{Vickers_X-ray_nitrides}{authors}{M.~A. Moram and M.~E. Vickers},
  \emph{Rep. Prog. Phys.}, {\bf 72}\penalty0 (3),  1  (2009).
\newblock \doi{10.1088/0034-4885/72/3/036502}.

\bibitem[Schroeder et~al.(2015)Schroeder, Ingason, Ros\'{e}n, and
  Thin]{Schroeder_Poor_MgO}
\nmbibLink{Schroeder_Poor_MgO}{authors}{J.L. Schroeder, A.S. Ingason,
  J.~Ros\'{e}n, and J.~Birch Thin}, \emph{J. Cryst Growth.}, {\bf 420},  22
  (2015).
\newblock \doi{10.1016/j.jcrysgro.2015.03.010}.

\bibitem[Vogt and Bierwagen(2016)]{Vogt2016a}
\nmbibLink{Vogt2016a}{authors}{P.~Vogt and O.~Bierwagen}, \emph{Appl. Phys.
  Lett.}, {\bf 109},  062103  (2016).
\newblock \doi{http://dx.doi.org/10.1063/1.4960633}.

\bibitem[Ahmed et~al.(1996)Ahmed, Sakai, Ota, Aoki, Ikemiya, and
  Hara]{Ahmed_MgOAnnealing}
\nmbibLink{Ahmed_MgOAnnealing}{authors}{F.~Ahmed, K.~Sakai, H.~Ota, R.~Aoki,
  N.~Ikemiya, and S.~Hara}, \emph{J. Low Temp. Phys.}, {\bf 105},  1343
  (1996).
\newblock \doi{10.1007/BF00753887}.

\bibitem[Dietz et~al.(1971)Dietz, Parisot, and Meixner]{Dietz_Raman}
\nmbibLink{Dietz_Raman}{authors}{R.~E. Dietz, G.~I. Parisot, and A.~E.
  Meixner}, \emph{Phys. Rev. B}, {\bf 4}\penalty0 (7),  2302  (1971).
\newblock \doi{10.1103/PhysRevB.4.2302}.

\bibitem[Mironova-Ulmane et~al.(2007)Mironova-Ulmane, Kuzmin, Steins, Grabis,
  Sildos, and P\"{a}rs]{Mironova_Raman}
\nmbibLink{Mironova_Raman}{authors}{N.~Mironova-Ulmane, A.~Kuzmin, I.~Steins,
  J.~Grabis, I.~Sildos, and M.~P\"{a}rs}, \emph{J. Phys. Conf. Ser.}, {\bf
  93}\penalty0 (012039)  (2007).
\newblock \doi{10.1088/1742-6596/93/1/012039}.

\bibitem[Arthur(2002)]{Arthur_MBE}
\nmbibLink{Arthur_MBE}{authors}{J.~R. Arthur}, \emph{Surf. Sci.}, {\bf 500},
  189  (2002).
\newblock \doi{10.1016/S0039-6028(01)01525-4}.

\bibitem[Tenne et~al.(2007)Tenne, Gonenli, Soukiassian, Schlom, Nakhmanson,
  Rabe, and Xi]{Tenne_SrTiO3}
\nmbibLink{Tenne_SrTiO3}{authors}{D.~A. Tenne, I.~E. Gonenli, A.~Soukiassian,
  D.~G. Schlom, S.~M. Nakhmanson, K.~M. Rabe, and X.~X. Xi}, \emph{Phys. Rev.
  B}, {\bf 76}\penalty0 (2),  024303  (2007).
\newblock \doi{10.1103/PhysRevB.76.024303}.

\bibitem[Aytan et~al.(2017)Aytan, Debnath, Kargar, Barlas, Lacerda, Li, Lake,
  Shi, and Balandin]{Aytan_2MSuppression}
\nmbibLink{Aytan_2MSuppression}{authors}{E.~Aytan, B.~Debnath, F.~Kargar,
  Y.~Barlas, M.~M. Lacerda, J.~X. Li, R.~K. Lake, J.~Shi, and A.~A. Balandin},
  \emph{Appl. Phys. Lett.}, {\bf 111}\penalty0 (25)  (2017).
\newblock \doi{10.1063/1.5009598}.

\bibitem[Betto et~al.(2017)Betto, Peng, Porter, Berti, Calloni, Ghiringhelli,
  and Brookes]{Betto_J}
\nmbibLink{Betto_J}{authors}{D.~Betto, Y.~Y. Peng, S.~B. Porter, G.~Berti,
  A.~Calloni, G.~Ghiringhelli, and N.~B. Brookes}, \emph{Phys. Rev. B}, {\bf
  96}\penalty0 (2)  (2017).
\newblock \doi{10.1103/PhysRevB.96.020409}.

\bibitem[Gandhi et~al.(2011)Gandhi, Huang, Yang, Chan, Cheng, Ma, and
  Wu]{Gandhi_Jvs2M}
\nmbibLink{Gandhi_Jvs2M}{authors}{A.~C. Gandhi, C.-Y. Huang, C.~C. Yang, T.~S.
  Chan, C.-L. Cheng, Y.-R. Ma, and S.~Y. Wu}, \emph{Nanoscale Res. Lett.}, {\bf
  6},  485  (2011).
\newblock \doi{10.1186/1556-276X-6-485}.

\bibitem[Aspnes et~al.(1979)Aspnes, Theeten, and Hottier]{Elli1}
\nmbibLink{Elli1}{authors}{D.~E. Aspnes, J.~B. Theeten, and F.~Hottier},
  \emph{Phys. Rev. B}, {\bf 20}\penalty0 (8),  3292  (1979).
\newblock \doi{10.1103/PhysRevB.20.3292}.

\bibitem[Ohl\'{i}dal et~al.(2017)Ohl\'{i}dal, Franta, and Necas]{Elli2}
\nmbibLink{Elli2}{authors}{I.~Ohl\'{i}dal, D.~Franta, and D.~Necas},
  \emph{Appl. Surf. Sci.}, {\bf 421},  687  (2017).
\newblock \doi{10.1016/j.apsusc.2016.10.186}.

\bibitem[Kang et~al.(2007)Kang, Lee, and Lee]{Opt_Kang}
\nmbibLink{Opt_Kang}{authors}{T.~D. Kang, H.~S. Lee, and H.~Lee}, \emph{J.
  Korean Phys. Soc.}, {\bf 50}\penalty0 (3),  632  (2007).
\newblock \doi{10.3938/jkps.50.632}.

\bibitem[R\"{o}dl and Bechstedt(2012)]{Opt_Roedl}
\nmbibLink{Opt_Roedl}{authors}{C.~R\"{o}dl and F.~Bechstedt}, \emph{Phys. Rev.
  B}, {\bf 86}\penalty0 (23),  235122  (2012).
\newblock \doi{10.1103/PhysRevB.86.235122}.

\bibitem[Ghosh et~al.(2015)Ghosh, Nelson, Abdallah, and Zollner]{Opt_Ghosh}
\nmbibLink{Opt_Ghosh}{authors}{A.~Ghosh, C.~M. Nelson, L.~C. Abdallah, and
  S.~Zollner}, \emph{J. Vac. Sci. Technol. A}, {\bf 33}\penalty0 (6),  061203
  (2015).
\newblock \doi{10.1116/1.4932514}.

\bibitem[Battiato et~al.(2016)Battiato, Giangregorio, Catalano, Losurdo, and
  Malandrino]{Opt_Battiato}
\nmbibLink{Opt_Battiato}{authors}{S.~Battiato, M.~M. Giangregorio, M.~R.
  Catalano, R.~Lo Nigro~M. Losurdo, and G.~Malandrino}, \emph{RSC Adv.}, {\bf
  6}\penalty0 (37),  30813  (2016).
\newblock \doi{10.1039/C6RA05510A}.

\bibitem[Kaspar et~al.(2016)Kaspar, Sushko, Heald, Papadogianni, Tschammer,
  Bierwagen, and Chambers]{Contact_Patterns}
\nmbibLink{Contact_Patterns}{authors}{T.~C. Kaspar, P.~V. Sushko, M.~E. Bowden
  S.~M. Heald, A.~Papadogianni, C.~Tschammer, O.~Bierwagen, and S.~A.
  Chambers}, \emph{Phys. Rev. B}, {\bf 94},  155409  (2016).
\newblock \doi{10.1103/PhysRevB.94.155409}.

\bibitem[Yamauchi et~al.(2015)Yamauchi, Hamasaki, Shibuya, Saito, Tsuchimine,
  Koyama, Matsuda, and Yoshimoto]{NiO_Al2O3}
\nmbibLink{NiO_Al2O3}{authors}{R.~Yamauchi, Y.~Hamasaki, T.~Shibuya, A.~Saito,
  N.~Tsuchimine, K.~Koyama, A.~Matsuda, and M.~Yoshimoto}, \emph{Sci. Rep},
  {\bf 5},  14385  (2015).
\newblock \doi{10.1038/srep14385}.

\end{thebibliography}

\end{document}